\documentclass[pdftex,dvipsnames,usenames,aps,superscriptaddress,twocolumn,amsmath,showkeys]{revtex4-1}

\usepackage{hyperref}
\usepackage{float}
\usepackage{dcolumn}
\usepackage{graphicx}
\usepackage{color}
\usepackage{amssymb}
\usepackage{bold-extra}
\usepackage{caption}
\usepackage{subcaption}

\def\figref#1{Fig.~\ref{fig:#1}}
\def\figlab#1{\label{fig:#1}}  
\def\eqref#1{Eq.~(\ref{eq:#1})}

\def\eqlab#1{\label{eq:#1}}
\newcommand*{\secref}[1]{Section~\ref{sec:#1}}
\newcommand*{\seclab}[1]{\label{sec:#1}}
\newcommand{\Omit}[1]{}

\def\KVI{KVI-Center for Advanced Radiation Technology, University Groningen, P.O. Box 72, 9700 AB Groningen, The Netherlands}
\def\AIVUB{Astrophysical Institute, Vrije Universiteit Brussel, Pleinlaan 2, 1050 Brussels, Belgium}
\def\VUB{Interuniversity Institute for High-Energy, Vrije Universiteit Brussel, Pleinlaan 2, 1050 Brussels, Belgium}
\def\NIKHEF{NIKHEF, Science Park Amsterdam, 1098 XG Amsterdam, The Netherlands}
\def\IMAPP{Department of Astrophysics/IMAPP, Radboud University Nijmegen, P.O. Box 9010,  6500 GL Nijmegen, The Netherlands}
\def\CWI{Center for Mathematics and Computer Science (CWI), PO Box 94079, 1090 GB Amsterdam, The Netherlands}
\def\TUe{Department of Applied Physics, Eindhoven University of Technology (TU/e), PO Box 513, 5600 MB Eindhoven, The Netherlands}
\def\ASTRON{Netherlands Institute of Radio Astronomy (ASTRON),  Postbus 2, 7990 AA Dwingeloo, The Netherlands}
\def\MPIB{Max-Planck-Institut f\"{u}r Radioastronomie, P.O. Box 20 24,  Bonn, Germany}
\def\UCI{Department of Physics and Astronomy, University of California Irvine, Irvine, CA 92697-4575, USA}
\def\DTU{DTU Space, National Space Institute, Technical University of Denmark, Elektrovej 328, 2800 Lyngby, Denmark}

\begin{document}

\title{Influence of Atmospheric Electric Fields on the Radio Emission from Extensive Air Showers.}

\author{T.~N.~G.~Trinh} \email[]{t.n.g.trinh@rug.nl}  \affiliation{\KVI}
\author{O.~Scholten}  \affiliation{\KVI}   \affiliation{\VUB}
\author{S.~Buitink} \affiliation{\AIVUB} \affiliation{\IMAPP}
\author{A.~M.~van den Berg}  \affiliation{\KVI}
\author{A.~Corstanje} \affiliation{\IMAPP}
\author{U.~Ebert} \affiliation{\CWI} \affiliation{\TUe}
\author{J.~E.~Enriquez} \affiliation{\IMAPP}
\author{H.~Falcke}  \affiliation{\IMAPP} \affiliation{\NIKHEF} \affiliation{\ASTRON} \affiliation{\MPIB}
\author{J.~R.~H\"orandel}  \affiliation{\IMAPP} \affiliation{\NIKHEF}
\author{C.~K\"{o}hn} \affiliation{\DTU}
\author{A.~Nelles}  \affiliation{\IMAPP} \affiliation{\UCI}
\author{J.~P.~Rachen} \affiliation{\IMAPP}
\author{L.~Rossetto}  \affiliation{\IMAPP}
\author{C.~Rutjes} \affiliation{\CWI}
\author{P.~Schellart} \affiliation{\IMAPP}
\author{S.~Thoudam} \affiliation{\IMAPP}
\author{S.~ter Veen} \affiliation{\IMAPP}
\author{K.D.~de Vries} \affiliation{\VUB}

\date{\today}

\begin{abstract}
The atmospheric electric fields in thunderclouds have been shown to significantly modify the intensity and polarization patterns of the radio footprint of cosmic-ray-induced extensive air showers.
Simulations indicated a very non-linear dependence of the signal strength in the frequency window of 30-80~MHz on the magnitude of the atmospheric electric field. In this work we present an explanation of this dependence based on Monte-Carlo simulations, supported by arguments based on electron dynamics in air showers and expressed in terms of a simplified model. We show that by extending the frequency window to lower frequencies additional sensitivity to the atmospheric electric field is obtained.
\end{abstract}


\maketitle

\section{Introduction}
When a high-energy cosmic-ray particle enters the upper layer of the atmosphere, it generates many secondary high-energy particles and forms a cosmic-ray-induced air shower. Since these particles move with velocities near the speed of light, they are concentrated in the thin shower front extending over a lateral distance of the order of 100 m, called the pancake. In this pancake the electrons and positrons form a plasma in which electric currents are induced. These induced currents emit electromagnetic radiation that is strong and coherent at radio-wave frequencies due to the length scales that are relevant for this process~\cite{Scholten:2009}.
Recent observations of radio-wave emission from cosmic-ray-induced extensive air showers~\cite{Falcke:2005, Codalema:2009, Aab:2014, Buitink:2013, Schellart:2014, Nelles:2015, Corstanje:2015} have shown that under fair-weather conditions there is a very good understanding of the emission mechanisms~\cite{Huege:2012}.
It is understood that there are two mechanisms for radio emission that determine most of the observed features. The most important contribution is due to an electric current, that is induced by the action of the Lorentz force when electrons and positrons move through the magnetic field of the Earth~\cite{Kahn:1966,Scholten:2008}.
The Lorentz force induces a transverse drift of the electrons and positrons in opposite directions such that they contribute coherently to a net transverse electric current in the direction of the Lorentz force $\bf{v}\times\bf{B}$ where $\bf{v}$ is the propagation velocity vector of the shower and $\bf{B}$ is the Earth's magnetic field. The radiation generated by the transverse current is polarized linearly in the direction of the induced current.  A secondary contribution results from the build up of a negative charge-excess in the shower front. This charge excess is due to the knock-out of electrons from air molecules by the shower particles, and gives rise to radio emission that is polarized in the radial direction to the shower axis~\cite{Askaryan:1962,Krijn:2010}.
The total emission observed at ground level is the coherent sum of both components. Because the two components are polarized in different directions, they are added constructively or destructively depending on the positions of the observer relative to the shower axis.
Since the particles move with relativistic velocities the emitted radio signal in air, a dielectric medium having non unity refractive index, is subject to relativistic time-compression effects. The radio pulse is therefore enhanced at the Cherenkov angle~\cite{Krijn:2011,Nelles:2015}.
Another consequence of the relativistic velocities is that the emission is strongly beamed and the radio emission is only visible in the footprint underneath the shower, limited to an area with a diameter of about 600 meters. As is well understood~\cite{Scholten:2008}, under fair-weather circumstances we see that the signal amplitude is proportional the energy of the cosmic ray and thus to the number of electrons and positrons in the extensive air shower~\cite{Buitink:2013}. We note that this proportionality of the radio emission to the number of electrons and positrons does no longer hold in the presence of strong atmospheric electric fields which is the main subject of this work.
The frequency content of the pulse is solely dependent on the geometry of the electric currents in the shower~\cite{deVries:2013}. As shown in the present work, the presence of strong atmospheric electric fields not only affects the magnitudes of the induced currents but, equally important, their spatial extent and thus the frequencies at which coherent radio waves are emitted.

There are several models proposed to describe radio emission from air showers: the macroscopic models MGMR~\cite{Scholten:2008}, EVA~\cite{Werner:2012} calculating the emission of the bulk of electrons and positrons described as currents; the microscopic models ZHAires~\cite{AlvarezMuñiz:2012}, CoREAS~\cite{Huege:2013} based on full Monte-Carlo simulation codes; and SELFAS2~\cite{Marin:2012}, a mix of macroscopic and microscopic approaches. All approaches agree in describing the radio emission~\cite{Huege:2013b}.

\begin{figure}[h]
                \includegraphics[width=0.48\textwidth]{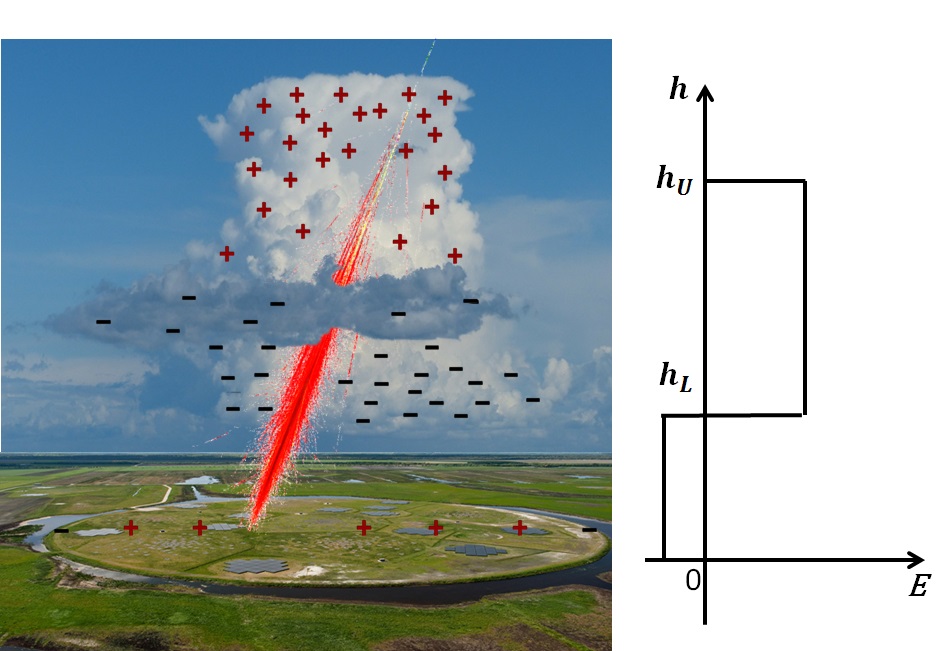}
               \caption{A schematic structure of a thundercloud is given where charge is accumulated at the bottom and the top layer. An air shower (in red) is passing through the thundercloud. The LOFAR core is seen as a circular structure on the ground where a few LOFAR antenna stations can be distinguished. The structure of the induced electric field is given schematically on the right hand side.}
                \figlab{Clouds}
\end{figure}

First measurements of the radio footprint of extensive air showers, made during periods where there were thunderstorms in the area, so-called thunderstorm conditions, have been reported by the LOPES~\cite{Buitink:2007, Apel:2011} collaboration. It was seen that the amplitude of the radiation was strongly affected by the atmospheric electric fields~\cite{Buitink:2010a}. More recently detailed measurements of the radio footprint, including its polarization were reported by the LOFAR~\cite{Schellart:2015} collaboration. The latter observations make use of the dense array of radio antennas near the core of the LOFAR radio telescope~\cite{Haarlem:2013}, a modern radio observatory designed for both astronomical and cosmic ray observations (see~\figref{Clouds}). At LOFAR two types of radio antennas are deployed where most cosmic ray observations have been made using the low-band antennas (LBA) operating in the 30~MHz to 80~MHz frequency window which is why we concentrate on this frequency interval in this work.
In the observations with LOFAR, made during thunderstorm conditions, strong distortions of the polarization direction as well as the intensity and the structure of the radio footprint were observed~\cite{Schellart:2015}. These events are called 'thunderstorm events' in this work.
The differences from fair-weather radio footprints of these thunderstorm events can be explained as the result of atmospheric electric fields and, in turn, can be used to probe the atmospheric electric fields~\cite{Schellart:2015}.

The effect of the atmospheric electric field on each of the two driving mechanisms of radio emission, transverse current and charge excess, depends on its orientation with respect to the shower axis. As we will show, the component parallel to the shower axis, $\bf{E_\parallel}$, increases the number of either electrons or positrons, depending on its polarity, and decreases the other. However, there is no evidence that this expected change in the charge excess is reflected in a change in the radio emission as can be measured with the LOFAR LBAs.
The component perpendicular to the shower axis, $\bf{E}_\perp$, does not affect the number of particles but changes the net transverse force acting on the particles. As a result the magnitude and the direction of the transverse current changes and thus the intensity and the polarization of the emitted radiation. However, simulations show that when increasing the atmospheric electric field strength up to ${E_\perp}=50$~kV/m, the intensity increases, as expected naively, after which the intensity starts to saturate.

In this work, we show that the influence of atmospheric electric fields can be understood from the dynamics of the electrons and positrons in the shower front as determined from Monte Carlo simulations using CORSIKA~\cite{Heck:1998}.
The electron dynamics is interpreted in a simplified model to sharpen the physical understanding of these findings.

\section{Radio-emission simulations}
\label{radioemission}
The central aim of this work is to develop a qualitative understanding of the dependence of the emitted radio intensity on the strength of the atmospheric electric field. For the simulation we use the code CoREAS~\cite{Huege:2013} which performs a microscopic calculation of the radio signal based on a Monte Carlo simulation of the air shower generated by CORSIKA~\cite{Heck:1998}.
The input parameters can be found in the Appendix. The particles in the shower are stored at an atmospheric depth of 500~g/cm$^2$, corresponding to a height of 5.7~km, near $X_\mathrm{max}$, the atmospheric depth where the number of shower particles is largest, for later investigation of the shower properties.
The radio signal is calculated at sea level as is appropriate for LOFAR.
The pulses are filtered by a 30~MHz to 80~MHz block band-pass filter corresponding to the LOFAR LBA frequency range. The total power is the sum of the amplitude squared over all time bins. The radiation footprints, representing the total power, (see \figref{Epara_LOFAR} and  \figref{Fperp_LOFAR}) are plotted in the shower plane, with axes in the direction of $\bf{v}\times\bf{B}$ and $\bf{v}\times\bf{v}\times\bf{B}$.

We have checked that proton induced showers show very similar features as presented here. We study iron showers to diminish effects from shower to shower fluctuations. Since these fluctuations are due to the stochastic nature of the first  high-energy interactions, they are larger in proton showers than in iron showers where there are many more nucleons involved in the initial collision. These fluctuations tend to complicate the interpretation of the numerical calculations since the changes observed in radio emission pattern can be due to these fluctuations or, more interestingly, to the effects of atmospheric electric fields.

\begin{figure}[htb]
        \centering
              \includegraphics[width=0.45\textwidth]{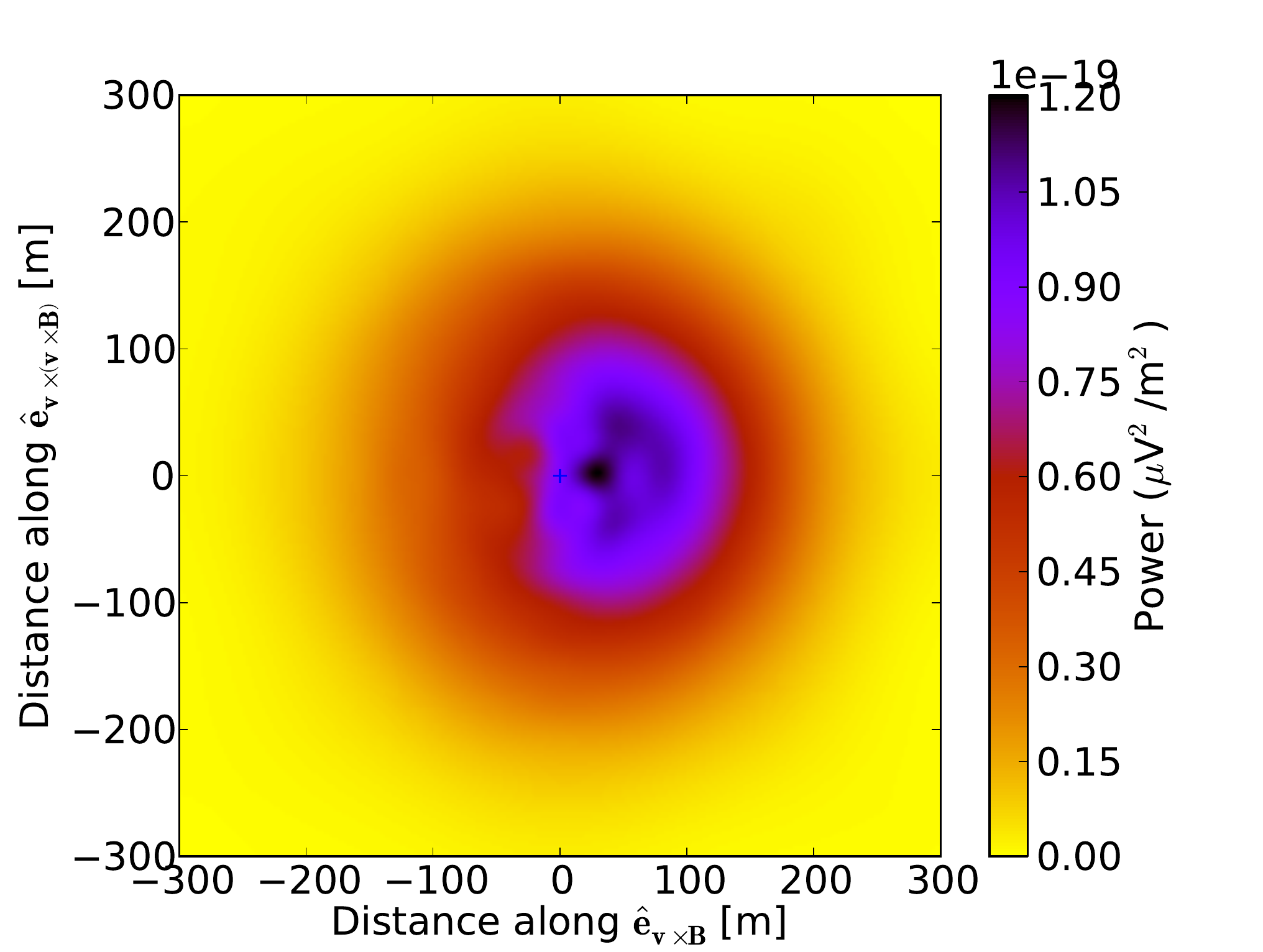}
              \includegraphics[width=0.45\textwidth]{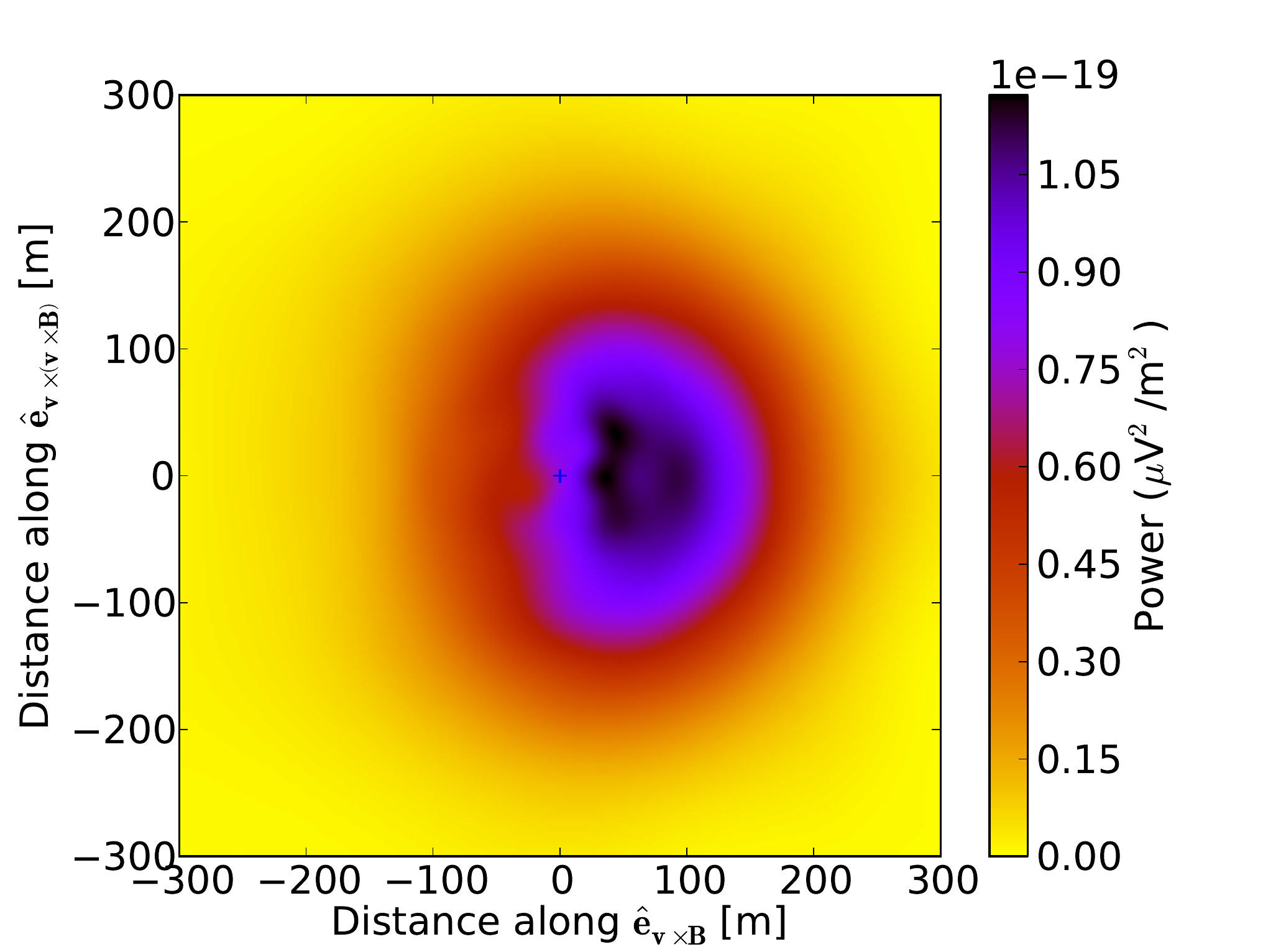}
               \includegraphics[width=0.45\textwidth]{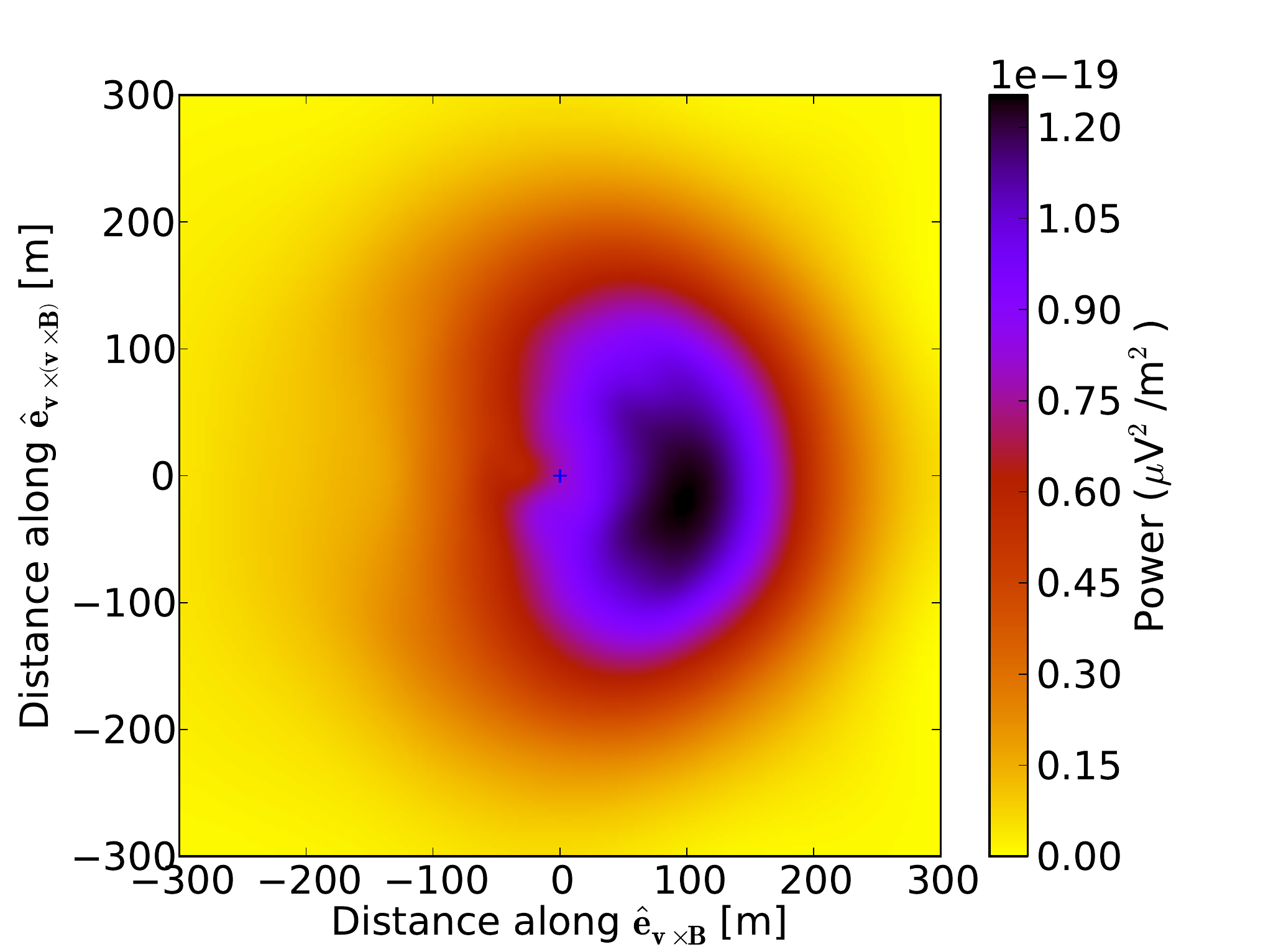}
    \caption{Intensity footprints of $10^{15}$~eV vertical showers for the 30 - 80~MHz band for the case of no electric field (top), $E_\parallel$ = 50~kV/m (middle), and $E_\parallel$ = 100~kV/m (bottom).}
\figlab{Epara_LOFAR}
\end{figure}

\begin{figure}[htb]
        \centering
                \includegraphics[width=0.45\textwidth]{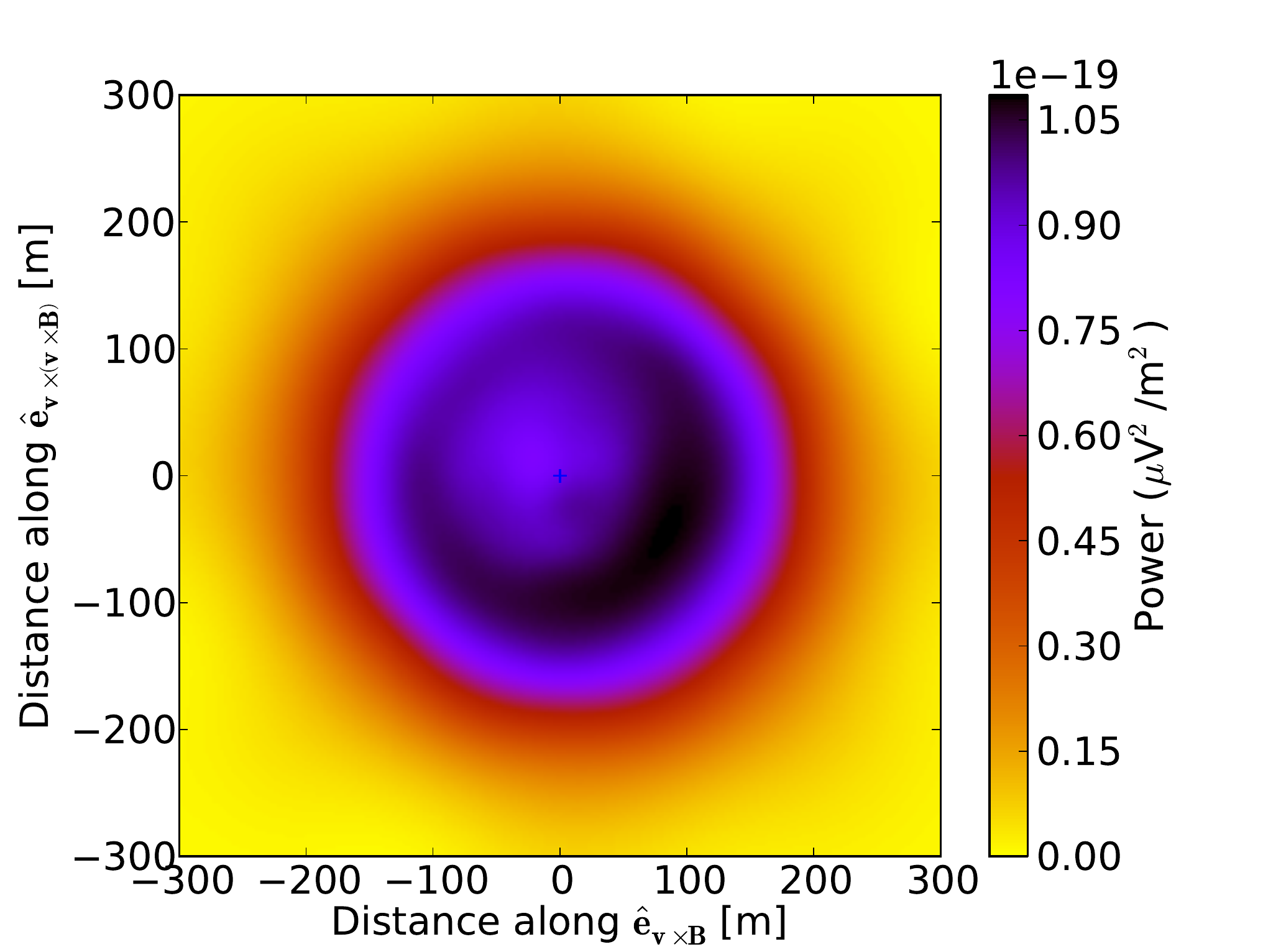}
                \includegraphics[width=0.45\textwidth]{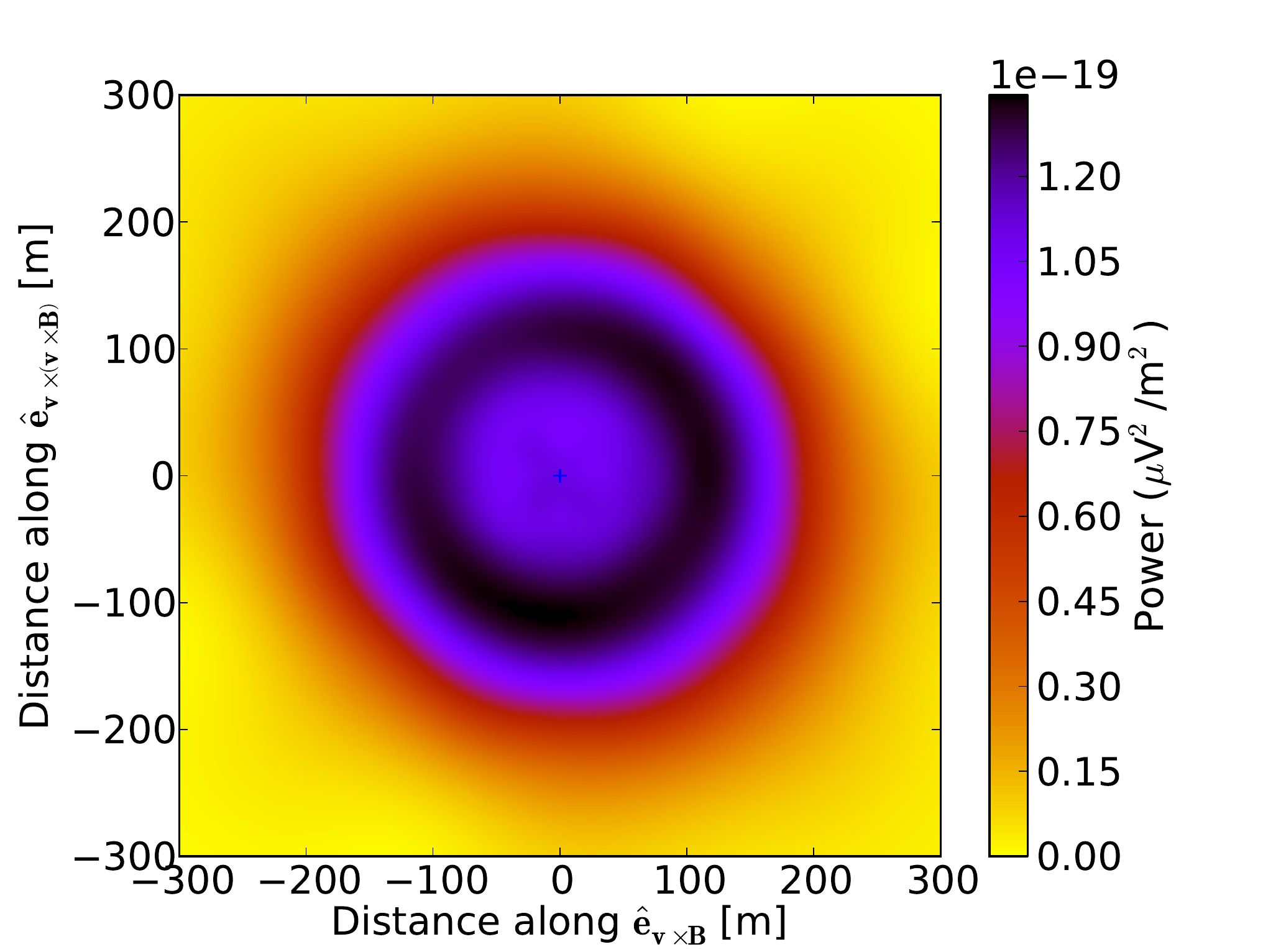}
    \caption{Intensity footprints of $10^{15}$~eV vertical showers for the 30 - 80~MHz band for the case of $F_\perp$ = 50~keV/m (top) and $F_\perp$ = 100~keV/m (bottom).}
\figlab{Fperp_LOFAR}
\end{figure}

As the aim of this paper is to obtain a deeper insight in the dependence of the radio-footprint of an extensive air shower on the strength of the electric fields, we have concentrated on one particular atmospheric field configuration that appears typical for at least half the events that are recorded under thunderstorm conditions.
We assume a two-layer electric-field configuration much like the one introduced in Ref.~\cite{Schellart:2015}.
This structure of the fields is schematically shown in \figref{Clouds}. Physically this field configuration can be thought to originate from a charge accumulation at the bottom and the top of a thunderstorm cloud.
The boundaries between the layers are set at $h_L=3$~km and $h_U=8$~km. The height of 3~km is typical for the lower charge layer in the Netherlands (see Fig. 1 in Ref.~\cite{Venema:2000} showing an ice containing cloud as an example). In thunderclouds, the upper charge layer would typically be above 8 km altitude. In this work, the height of 8~km is chosen since we are not sensitive to even higher altitudes where there are few air-shower particles~\cite{Schellart:2015}.
The strength of the field in the lower layer is fixed at a certain fraction of the value of the field in the upper layer ranging from $h_L$ till $h_U$, oriented in the opposite direction.
The orientation of the electric field is not necessarily vertical as it depends on the orientation of the charge layers.
Finding the orientation of the field is thus an important challenge for the actual measurement. 
As we will show in the following sections the sensitivity of the radio footprint is rather different for fields parallel and perpendicular to the direction of cosmic ray. To show this we study these two cases separately to have a discussion of these sensitivities as clean as possible. This may give rise to an unphysical field structure in some cases (see Sec. \ref{Eperp}). To obtain a physical field configuration with vanishing curl one could have added a parallel component where the magnitude depends on the assumed orientation of the charge layer. We have opted not to introduce this arbitrariness since the sensitivity to the parallel electric field is small.
In this work we consider field strengths in the upper layer of up to 100~kV/m which is below the runaway breakdown limit of 284 kV/m at sea level~\cite{Dwyer:2003, Marshall:2005} and of 110~kV/m at 8~km. Balloon observations show that the electric fields vary with altitude~\cite{Stolzenburg:2007}. The electric fields used in the simulations are homogeneous within each layer and should be considered some average field. In \secref{adad} we argue that due to intrinsic inertia in the shower development the field effects are necessarily averaged over distances of the order of 0.5~km.
The change of orientation of the electric field at the height $h_L$ introduces a destructive interference between the radio emission of air showers in the two layers, generating a ring-like structure in the radio footprint.
Electric fields in thunderstorm conditions can be more complicated than the simple structure assumed here which will reflect in more intricate radio-footprints (see Ref.~\cite{Schellart:2015} for an example). These more complicated configurations will be the subject of a forth coming article.
It should be noted that the insight in the particle motion at the air-shower front under the influence of electric fields presented in this work is independent of the detailed structure of the field configuration.

\subsection{Parallel electric field}

\begin{figure}[h]
                \includegraphics[width=0.48\textwidth]{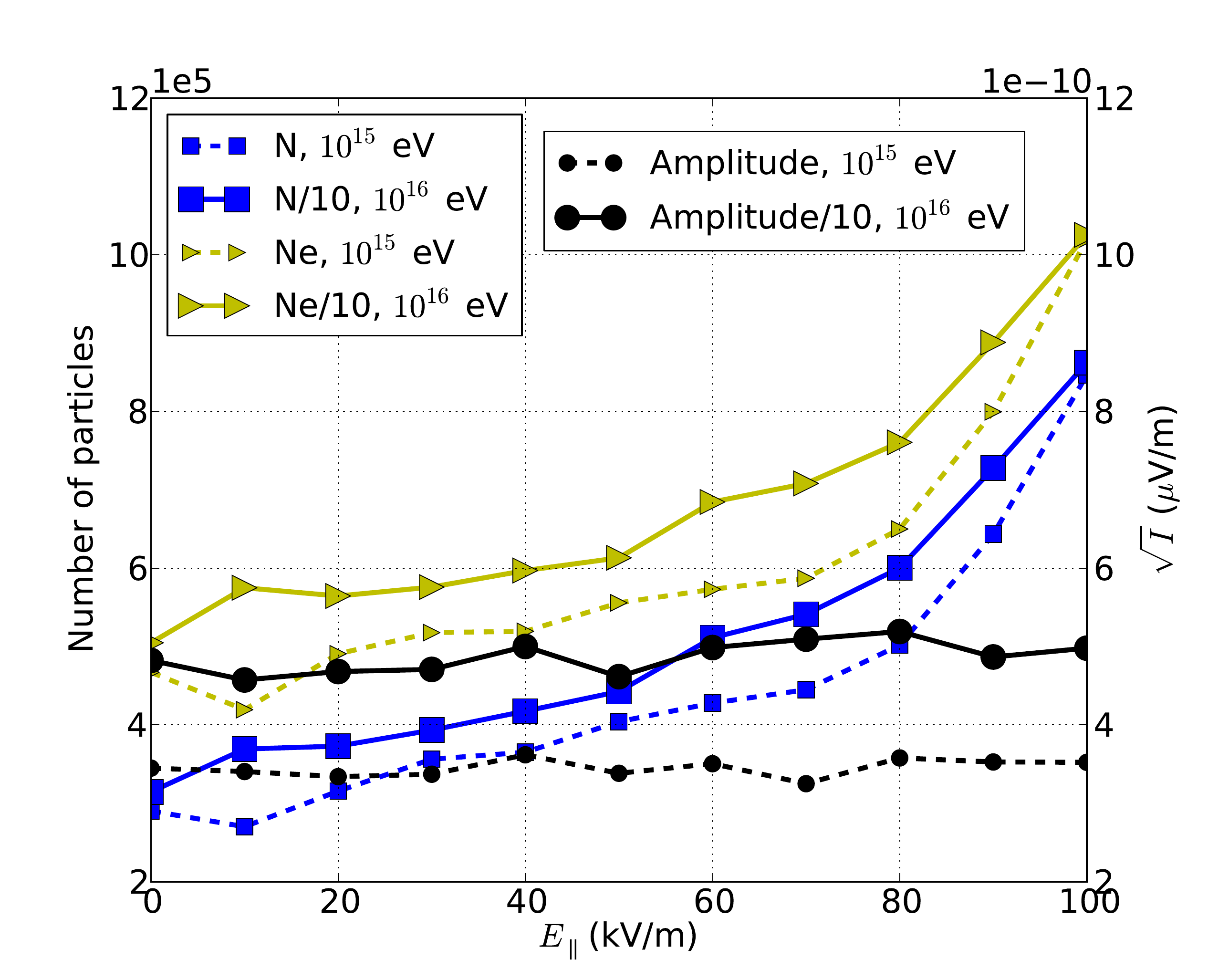}
               \caption{The number of electrons and positrons at the shower maximum (blue left axis), the number of electrons at the shower maximum (yellow left axis) and the maximum $\sqrt{I}$ (black right axis) for vertical $10^{15}$~eV showers (dashed lines)  and for vertical 10$^{16}$ eV showers (right solid lines) as a function of the parallel electric fields. For the 10$^{16}$ eV showers the number of particles and the pulse-amplitude are scaled down by a factor 10.}
                \figlab{power_Epara}
\end{figure}

To study the effects of a parallel electric field, the strength of the field in the upper layer is taken in the direction of the shower and is varied from 0 to 100~kV/m in steps of 10~kV/m. The upper-layer field points upward along the shower axis and accelerates electrons downward. The field in the lower layer is set at $k=0.3$ times the value in the upper layer, pointing in opposite direction. For simplicity we consider vertical showers.
As can be seen from \figref{power_Epara} the number of electrons and positrons at the shower maximum increases with increasing electric field while the square root of the power $\sqrt{I}$ in the radio pulse remains almost constant. The number of electrons also increases with increasing electric field. The number of positrons, not shown here, lightly decreases. Since there are fewer positrons than electrons the total number of electrons and positrons still increases.
Thus, for coherent emission, where the amplitude of the signal is proportional to the total number of electrons and positrons, one expects the signal strength to increase proportional to the total number of particles with the electric field. The simulation results in \figref{power_Epara} show clearly that, contrary to this expectation, the square root of the power $\sqrt{I}$ is almost constant. These features will be explained in \secref{E-par}. The fluctuations in the $\sqrt{I}$ are due to shower-to-shower fluctuations. The difference in the $\sqrt{I}$ (after scaling) between the $10^{15}$~eV shower and the $10^{16}$~eV shower appears due to slight difference in $X_{max}$.

As explained in \secref{E-par} the observed limited dependence is due to the fact that the additional low-energy electrons in the shower trail behind the shower front at a relatively large distance and thus do not contribute to coherent emission at the observed frequencies. The trailing behind the shower front of the low-energy electrons was also shown in Ref.~\cite{Buitink:2010}, but for the breakdown region.

Not only the strength of the signal, but also the structure of the radio footprints for the LOFAR LBA frequency range, as shown in \figref{Epara_LOFAR}, does not really depend on the strength of the parallel electric field. Furthermore, the bean shape (see the top panel of \figref{Epara_LOFAR}), typically observed in air showers in fair-weather condition, is also present in these footprints because the parallel electric fields have small effects on both the transverse-current and the charge-excess components.

\subsection{Transverse electric field}
\label{Eperp}

To investigate the effect of a transverse electric field we will take a simple geometry with a vertical shower and horizontally oriented electric fields. As mentioned in Sec. \ref{radioemission}, this does give rise to the situation where at height $h_L$ the horizontal component changes sign giving rise to a finite value for curl(E) which is not physical. This should, however, be regarded as the limiting result of the case where the cosmic ray is incident at a finite zenith angle crossing an almost horizontal charge layer. At the charge layer the direction of the electric field changes, i.e.\ the components transverse as well as longitudinal to the shower direction. Here we concentrate on the transverse component.

\begin{figure}[htb]
                \includegraphics[width=0.48\textwidth]{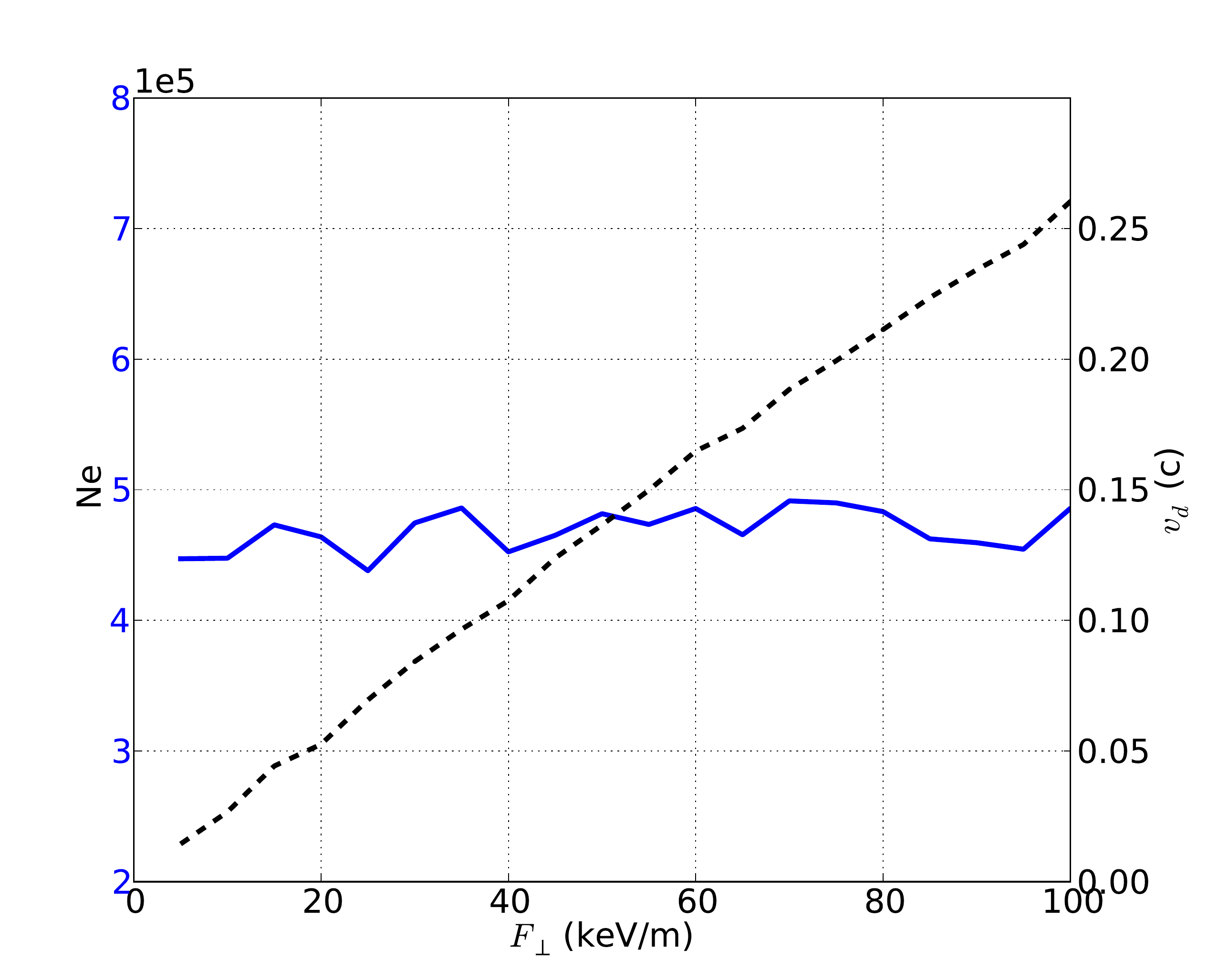}
               \caption{The number of electrons (solid blue line, left axis) and their drift velocity at $X_{max}$ (dashed black line, right axis) of vertical $10^{15}$~eV showers as a function of the net-transverse forces.}
                \figlab{vd_Fperp}
\end{figure}

The transverse electric field does not change the number of electrons, but instead increases the magnitude and changes the direction of the drift velocity of the electrons. This is shown in \figref{vd_Fperp} where the results of simulations are shown for vertical $10^{15}$~eV showers for the case in which the net force on the electrons (the sum of the Lorentz and the atmospheric electric field) is oriented transversely to the shower at an angle of 45$^\circ$ to the $\bf{v}\times\bf{B}$ direction.
The strength of the net-transverse force in the upper layer is varied from 5~keV/m to 100~keV/m in steps of 5~keV/m. For the lower level the electric field is chosen such that the net force acting on the electrons is a fraction 0.3 of that in the upper layer.
One observes that the number of electrons at the shower maximum stays rather constant while the transverse-drift velocity of these electrons increases almost linearly with the strength of the net force. The induced transverse current thus increases linearly with the net force.

Due to a strong increase in the transverse current contribution while the charge excess contribution remains constant, the asymmetry in the pattern diminishes and the radio footprint attains a better circular symmetry around the shower core.  The interference between radio emission in two layers introduces a destructive interference near the core which results in a ring-like structure in the intensity footprint which can clearly be distinguished in \figref{Fperp_LOFAR}. At the ring the signal reaches the maximum value.
For the case of $F_\perp$ = 50~keV/m (top panel of \figref{Fperp_LOFAR}), there is an asymmetry along the 45 $^{\circ}$ axis which is the direction of the net force $\bf{F_\perp}$ and results from the interference with the charge-excess component. For $F_\perp$ = 100~keV/m this asymmetry is even smaller.

\begin{figure}[htb]
                \includegraphics[width=0.48\textwidth]{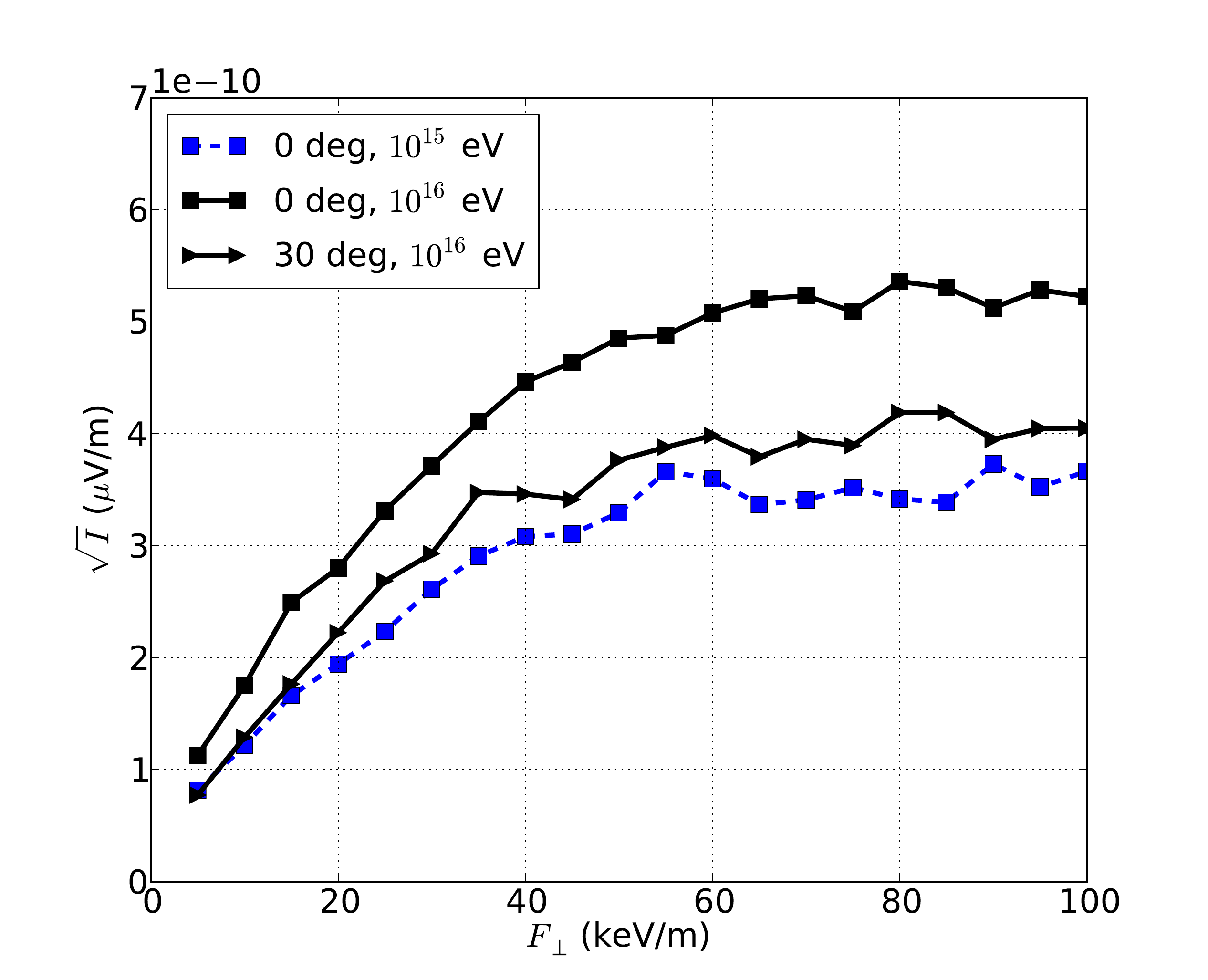}
               \caption{The square root of the power $\sqrt{I}$ at the ring of maximal intensity for vertical $10^{15}$~eV showers (dashed line) as well as for vertical and inclined $10^{16}$~eV showers (solid lines) as a function of the net-transverse force. For the 10$^{16}$ eV showers the $\sqrt{I}$ is scaled down by a factor 10.}
                \figlab{power_Fperp}
\end{figure}

Interestingly, \figref{power_Fperp} shows that the square root of the power $\sqrt{I}$ is proportional to the net force until about 50~keV/m where it starts to saturate. This appears to be a general feature, independently of shower geometry.
As we will argue in \secref{E-perp}, the saturation of the $\sqrt{I}$ is due to loss of coherence since with increasing transverse electric field the electrons trail at larger distances behind the shower front.

\section{Interpretation}
The radio emission simulation results clearly show that the strength of the radio signal saturates as a function of the applied transverse electric field and seems to be insensitive to the parallel component of the electric field. In this section we explain these observations on the basis of the electron dynamics as can be distilled from Monte Carlo simulations using CORSIKA. To interpret these we will use a simplified picture for the motion of the electrons behind the shower front.
Since effects of electric fields on electrons and positrons are almost the same but opposite in direction, we will, to simplify the discussion, concentrate on the motion of electrons.

The central point in the arguments presented here is the fact that the emitted radio-frequency radiation is coherent. The intensity of coherent radiation is proportional to the square of the number of particles while for incoherent radiation it is only linearly proportional. Since the number of particles is large, many tens of thousands, this is an important factor. To reach coherence the retarded distance between the particles should be small compared to the wavelength of the radiation.
For the present cases most of the emission is in near forward angles from the particle cascade which implies that the important length scale is the distance the electrons trail behind the shower-front~\cite{Scholten:2009}. When this distance is typically less than half a wavelength the electrons contribute coherently to the emitted radiation.
For the LOFAR frequency window of 30-80~MHz we assume that this coherence length to be 3~m. The challenge is thus to understand the distance the electrons trail behind the shower front.

In the discussion in this paper we distinguish a transverse force acting on the electrons and a parallel electric field. The transverse force is the (vectorial) sum of the Lorentz force derived from the magnetic field of the Earth and the force due to a transverse electric field. We limit our analysis to the electrons having an energy larger than 3~MeV since lower energetic ones contribute very little to radio emission.
\subsection{Energy-loss time of electrons}

\Omit{\begin{figure}[h]
        \centering
        \begin{subfigure}[h]{0.45\textwidth}
                \includegraphics[width=\textwidth]{cumulative_distance_Epara}
               \caption{E-parallel} \figlab{CumDist-para}
        \end{subfigure}
        \begin{subfigure}[h]{0.45\textwidth}
                \includegraphics[width=\textwidth]{cumulative_distance_Fperp}
                   \caption{E-perp} \figlab{CumDist-perp}
        \end{subfigure}
      \caption{The cumulative distance behind the shower front.}
\figlab{CumDist}
\end{figure}
}
In the simplified picture we use for the interpretation of the Monte-Carlo results we will assume that the energetic electrons are created at the shower front with a relatively small and randomly oriented transverse momentum component. Like the nomenclature used for the electric field, transverse implies transverse to the shower axis which is in fact parallel to the shower front.
After being created the energetic electron is subject to soft and hard collisions with air molecules through which it will loose energy. For high-energy electrons the Bremsstrahlung process is important, through which they may loose about half their energy.
The radiation length for high-energy electrons in air, mostly due to Bremsstrahlung, is $X_0 \sim 36.7$ g/cm$^{2}$~\cite{Amsler:2008} and their fractional energy loss per unit of atmospheric depth is $a = 1/X_0 = 0.0273$~cm$^{2}$/g~\cite{ESTAR}. In the low-energy regime, for energies larger than 3~MeV, soft collisions with air molecules take over which hardly depend on the initial energy of the electrons and the energy loss per  unit of atmospheric depth is almost constant $b = 1.67$~MeV\,cm$^2$/g~\cite{ESTAR}. The energy loss for low-energy as well as high-energy electrons can thus be parameterized as
\begin{equation}
-\frac{dU}{dX}=a\,U+b\,,
\end{equation}
where $U$ is the energy of electrons and $X$ is the atmospheric depth.

The distance $L$ (in [g/cm$^2$]) over which the electron energy is reduced to a fraction $1/\xi$ of the original energy, i.e.\ they loose an energy of $\Delta U=(1-1/\xi)U$, is thus
\begin{equation}
L_\xi= \frac{1}{a}\ln \left[\frac{a\,U_0+b}{a\,U_0/\xi+b} \right].
\end{equation}
Since these particles move with a velocity near the speed of light, where we use natural units $c = 1$, this corresponds to an energy-loss time $\tau_\xi$, given by
\begin{equation}
\tau_\xi(U) = \frac{L}{\rho} = \frac{1}{a\,\rho}\ln \left[ \frac{a\,U_0+b}{a\,U_0/\xi+b} \right] \;. \eqlab{taux}
\end{equation}
The air density $\rho$ is approximately~\cite{Marshall:1995}
\begin{equation}
\rho(z) = 1.208\cdot 10^{-3} \exp(z/8.4)\;\text{g/cm$^3$}\,,\figlab{rho}
\end{equation}
where $z$ is the altitude in km.

In our picture, developed to visualize the results from the full-scale Monte Carlo simulations, the energy-loss time plays a central role since it is the amount of time over which we will follow the particles after they are created at the shower front.
In our picture this thus plays the role of a life-time of the electrons after which they are assumed to have disappeared and may have re-appeared as a lower energy electron or have been absorbed by an air molecule.
In this energy-loss time, $\tau(U)$, thus several more complicating effects have been combined such as:
\begin{itemize}
\item In reality electrons of energy $U$ are created by more energetic particles. They are already  trailing some distance behind the front. This additional distance is absorbed in the definition of $\tau(U)$.
\item Once an electron is created at a certain energy we do not take its energy loss into account in calculating the properties of the shower front. Such a formulation can only be applied within a limited range of lost energy.
\end{itemize}
We will take both effects into account by making an appropriate choice for the parameter $\xi$. One consequence of the parametrization, which can be tested directly with simulations, is that the median distance  from the shower front of electrons (see \eqref{ADtau}) does not depend on the primary energy of showers. As shown in \figref{def_tau} this is obeyed rather accurately and furthermore the energy-dependence of the median distance behind the shower front, calculated using the approach discussed in the following section, can be fitted reasonably well by taking $\xi=e$.
We thus define
\begin{equation}
\tau(U) = \frac{L}{\rho} = \frac{1}{a\,\rho}\ln \left[ \frac{a\,U_0+b}{a\,U_0/e+b} \right] \;. \eqlab{tau}
\end{equation}

\begin{figure}[h]
                \includegraphics[width=0.48\textwidth]{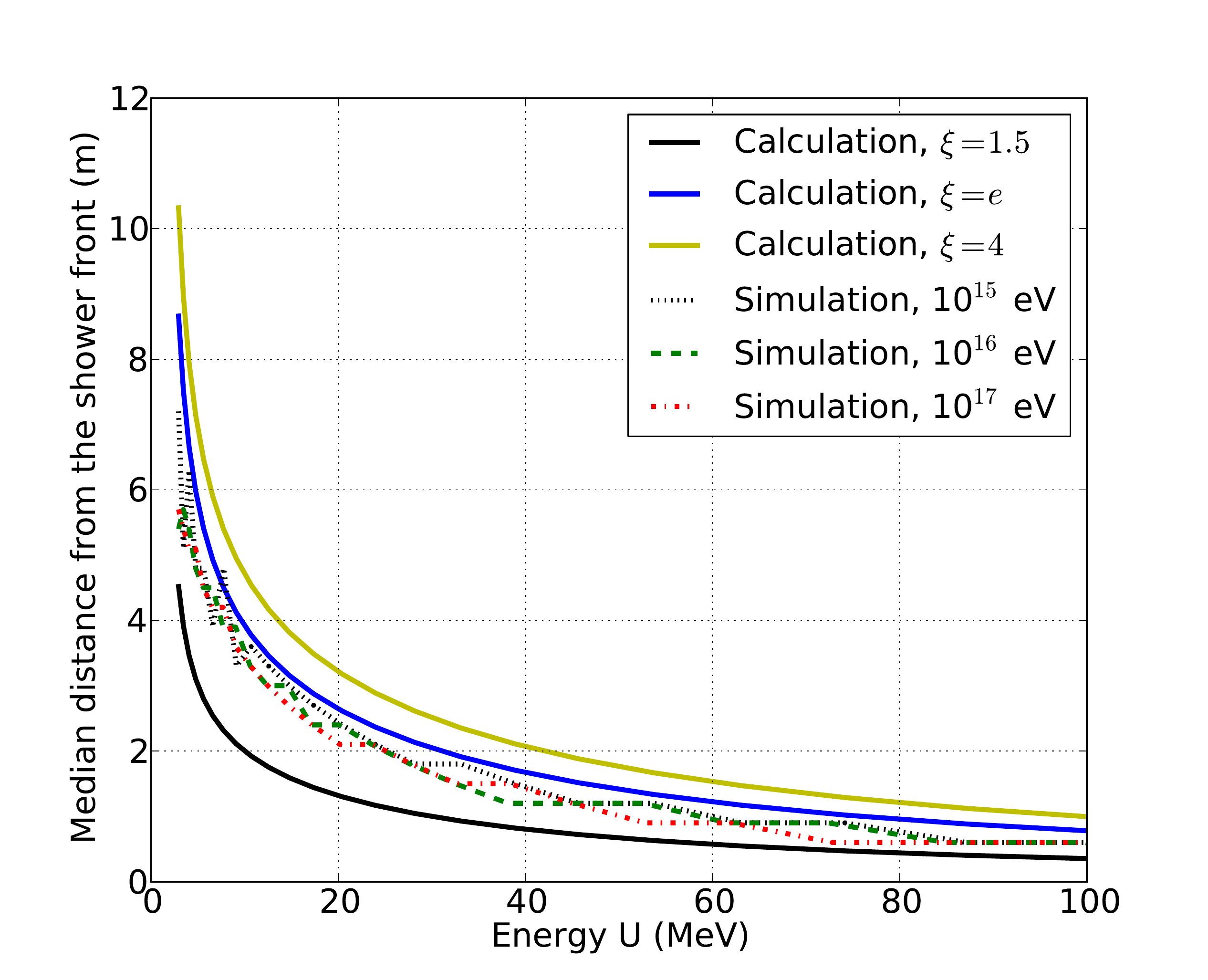}
               \caption{The median distance from the shower front of electrons for vertical $10^{15}$~eV,  10$^{16}$ eV and 10$^{17}$ eV showers in the absence of electric fields. The simple prediction based on \eqref{tau} and \eqref{ADtau} is also shown.}
                \figlab{def_tau}
\end{figure}

\subsection{Trailing distance}

As written before we assume that the electrons are generated at the shower front and from there drift to larger distances. Assuming the electrons disappear from the process after a time $\tau$, the electrons are generated at the shower front at a rate
\begin{equation}
\frac{dP(U)}{dt}=P_0( U)\frac{1}{\tau} \;\;;
\eqlab{generationrate}
\end{equation}
where $P_0(U)$ is the energy distribution at the shower maximum, as for example given by \cite{Nerling:2006}
\begin{equation}
P_0(U)= \frac{dN}{d( \ln \frac{U}{1\text{MeV}})}=\frac{U_0U}{(U+U_1)(U+U_2)}\,,
\eqlab{energy_spec}
\end{equation}
where $U_0, U_1, U_2$ are parameters that are determined from a Monte-Carlo simulation. The creation rate, \eqref{generationrate}, has been chosen such that the energy distribution in the shower pancake agrees with the observations
\begin{equation}
P(U)=\int_0^{ \tau} \frac{dP(U)}{dt}\,dt = P_0(U)\,.
\eqlab{NU}
\end{equation}
The parameters in \eqref{energy_spec} are chosen to reproduce the electron spectrum from the simulation at $10^{15}$~eV giving
$U_0=10^{6}$~MeV, $U_1=4.11$~MeV, and $U_2=105$~MeV.

To obtain an estimate of the distance the electrons trail behind the shower front we calculate the difference in forward velocity of the front and the electrons.
The air-shower front moves with the speed of light, while the electrons (mass $m_0$, energy U, random transverse momentum $P_\perp$ and longitudinal momentum $P_\parallel$) travel with a longitudinal velocity $v_\parallel$ which is less than $c$, given by
\begin{equation}
v_\parallel = \frac{P_\parallel}{U} =\frac{\sqrt{U^2-m_0^2-P_\perp^2}}{U}
\approx 1-\frac{m_0^2+P_\perp^2}{2\,U^2}\;,
\eqlab{vpara}
\end{equation}
where $\left(m_0^2+(P_\perp)^2\right)/U^2$ is assumed to be much smaller than $1$. After a time $t$, due to the velocity difference between the shower front and the electrons, the electrons are trailing behind the shower front by a distance
\begin{equation}
l = \int_0^t(c-v_\parallel)dt\,.
\end{equation}
Using the assumption that the energy of the electrons does not change significantly during the energy-loss time, and taking in addition $P_\perp$ to be time-independent, one obtains
\begin{equation}
l =  \frac{1}{2}\frac{\left(m_0^2+P_\perp^2\right)}{U^2}t\,,
\eqlab{delta}
\end{equation}
Based on the results of CORSIKA simulations we can introduce an effective mass of the electrons that includes the stochastic component of the transverse momentum, which is parameterized as
\begin{equation}
m_\perp^2=\langle m_0^2 + P_\perp^2 \rangle= m_0^2\left[1+ 10\,\left(U/m_0\right)^{2/3}\right] \;. \eqlab{PPerpAve}
\end{equation}
The median distance by which an electron can trail behind the shower front within its energy-loss time $\tau$, $D$, can now be written as
\begin{equation}
D = l(\tau/2) = \frac{m_\perp^2\tau}{4\,U^2} \,
\eqlab{ADtau}
\end{equation}
and is shown in \figref{def_tau}.

In the Monte Carlo simulation results shown in \figref{def_tau} as well as those discussed in the following sections the trailing distances are calculated with respect to a flat shower front thus ignoring the fact that in reality the shower front is curved. It has been checked that the corrections due to this curvature effect for distances up to 100~m from the shower core do not exceed 30~cm and thus can be ignored. In addition, electrons at very low energies do not contribute significantly to radio emission. Therefore, we have limited the analysis to the electrons having energies larger than 3~MeV and a distance to the shower core of less than 100~m.

\subsection{Influence of $\bf{E_\parallel}$}\seclab{E-par}

\begin{figure}[h]
                 \centering
                \includegraphics[width=0.45\textwidth]{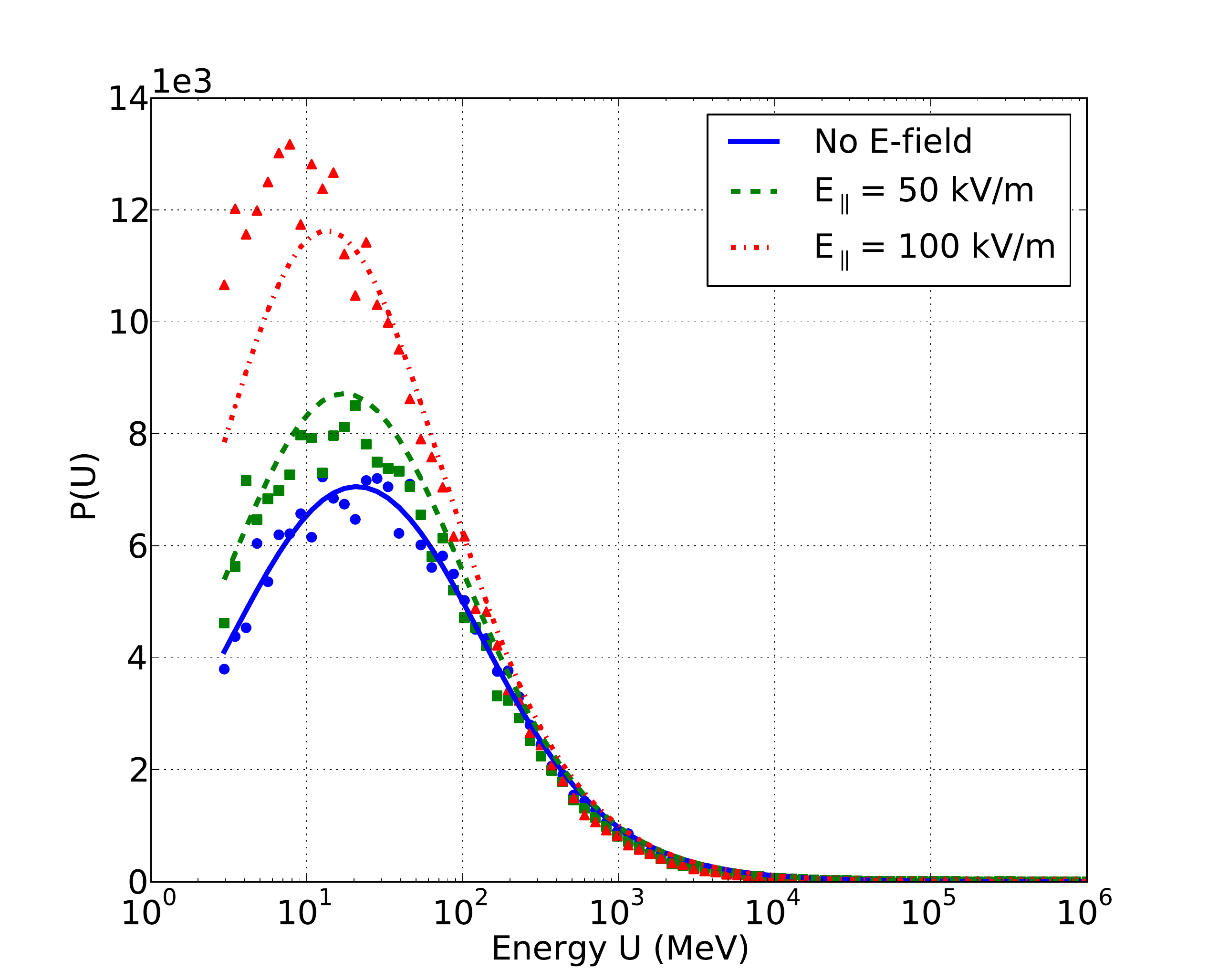}
               \caption{The number of electrons is shown as function of their energy for electric fields $E = 0$, $\bf{E_\parallel} = 50$ kV/m and $\bf{E_\parallel} = 100$ kV/m as obtained from the analytical calculations (solid and dashed curves) and from full CORSIKA simulations (markers) at $X = 500$ g/cm$^2$.}
            \figlab{E_dis}
\end{figure}

When a parallel electric field $\bf{E_\parallel}$ is applied to accelerate electrons downward, the CORSIKA results show a strong increase in the number of electrons (see \figref{E_dis}) as well as in the trailing distance behind the shower front (see \figref{mean_distance_Epara}). To understand these trends in our simple picture we first investigate the effect of the applied electric field on the energy-loss time.

\begin{figure}[h]
                 \centering
                \includegraphics[width=0.45\textwidth]{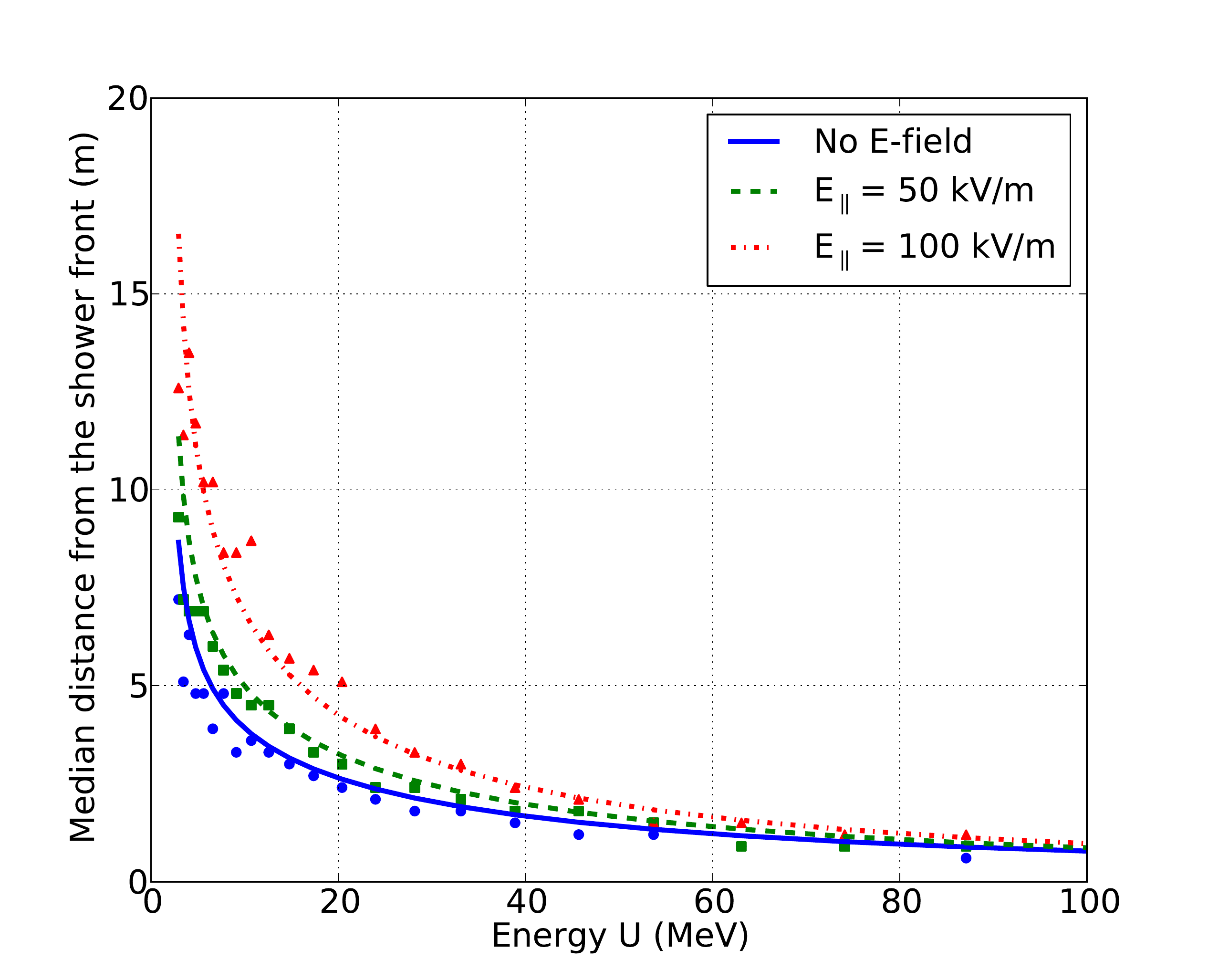}
               \caption{The distance by which the electrons travel in the absence and in the presence of the  parallel electric fields from analytical calculations (solid and dashed curves) and from CORSIKA simulations (markers).  }
            \figlab{mean_distance_Epara}
\end{figure}

When a parallel electric field $\bf{E_\parallel}$ is applied to accelerate electrons downward, the electrons gain energy from the electric field. Therefore, the energy-loss equation for the electrons is modified to
\begin{equation}
-\frac{dU}{dX}=a\,U+b - F_E/\rho \,, \eqlab{dUdxE}
\end{equation}
where $F_E = e\,E$. Since the field strengths we will consider are below the break-even value $E_{be}(z)= 1.67\cdot10^6\,\rho(z)$ V/cm~\cite{Marshall:1995} the electrons always loose energy and the r.h.s.\ of \eqref{dUdxE} is always positive. Stated more explicitly, we are not modeling the runaway breakdown process and we have therefore limited the electric field to strengths below the breakdown value of 100~kV/m at an altitude of 5.7 km.

The energy-loss time of the electrons inside the electric field is now modified to
\begin{equation}
\tau_E = \frac{L_E}{\rho} = \frac{1}{a\,\rho}\ln \left[ \frac{a\,U_0+b-F_E/\rho}{a\,U_0/e+b-F_E/\rho} \right] \;. \eqlab{tau-E}
\end{equation}

\begin{figure}[h]
                \includegraphics[width=0.48\textwidth]{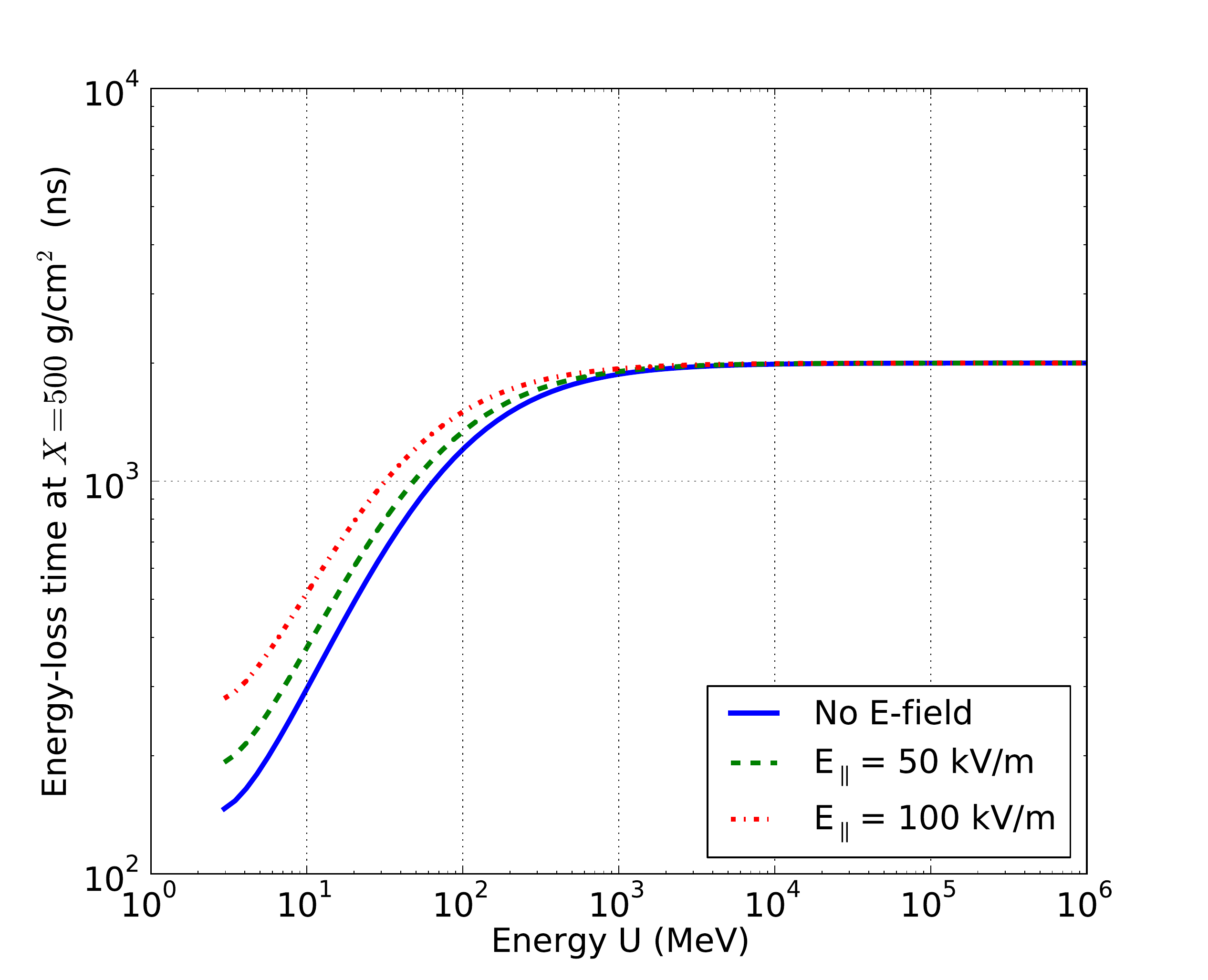}
               \caption{The energy-loss time of electrons at $X = 500$ g/cm$^2$ (at $X_{max}$) in the absence and in the presence of parallel electric fields.}
                \figlab{lifetime}
\end{figure}
\figref{lifetime} shows that the energy-loss time of low-energy electrons is larger in the presence of the electric field than in the absence of it. Inside the electric field, the low-energy electrons gain energy from the field, it thus takes longer for their energy to drop below the fraction $1/\xi$ of their original energy, and thus, according to our definition, live longer. In the high-energy regime, the electrons also gain energy from the electric field, but this gained energy is small compared to their own energy. As a result, the energy-loss time of high-energy electrons is almost unchanged.

The generation rate of electrons at the shower front will remain unchanged and is given by \eqref{generationrate}. The number of electrons in the shower pancake changes from \eqref{NU} to
\begin{equation}
P_E(U) = \int_0^{\tau_E} \frac{dP(U)}{dt}\,dt = P_0(U)\frac{\tau_E}{\tau}\;.
\label{NE_E}
\end{equation}
Since in the low-energy regime $\tau_E > \tau$, there is an increase in the number of low-energy electrons. In the high-energy regime, on the other hand, where $\tau_E\approx \tau$, the number of electrons is almost unchanged. \figref{E_dis} shows that this simple calculation can reproduce the main features shown in the CORSIKA simulations rather well except a small disagreement at the low-energy region in the presence of the electric field of 100~kV/m.
The presently considered field accelerates the electrons and thus decelerates the positrons. As a consequence the positron energy-loss time decreases resulting in a rather large decrease of the number of low-energy positrons while the number of high-energy positrons stays almost constant.

When the electron is subject to an electric field the energy-loss time is given by \eqref{tau-E} instead of by \eqref{tau} and as a result the expression for the median trailing distance \eqref{ADtau}, changes to
\begin{equation}
D(E_\parallel) = \frac{m_\perp^2\tau_E}{4\,U^2}\;,
\end{equation}
where the mean transverse momentum is given by \eqref{PPerpAve}.
The effect of the electric field on the median trailing distance is shown in \figref{mean_distance_Epara}. Due to the influence of electric field the low-energy electrons trail much further behind the shower front. The median trailing distance as calculated from CORSIKA simulations is also displayed in \figref{mean_distance_Epara}.
It shows that our simplified picture correctly explains the trends seen in the full Monte-Carlo simulations using the CORSIKA simulations.  In the absence of an electric field the trailing distance sharply decreases with increasing energy.
When applying an electric field that accelerates the electrons the Monte Carlo simulations show a considerable increase in the trailing distance behind the shower front. From the present simple picture this can be understood to be generated by the increased energy-loss time that generates a considerable trailing of electrons with energies below 50~MeV.

The interesting aspect for radio emission is the number of particles within a distance of typically half a wavelength of the shower front. For the LOFAR LBA frequency range this corresponds to a distance of about 3~m. Using \eqref{delta} as well as \eqref{generationrate} the distribution of the electrons over distance $l$ behind the shower front is given by
\begin{equation}
\frac{dP(U)}{dl} =\frac{dP(U)}{dt}\frac{dt}{dl}= \frac{P_0(U)}{\tau} \frac{2U^2}{\left(m_0^2+P_\perp^2\right)}\,.
\end{equation}
Therefore, the number of electrons within the distance $\Delta$ is
\begin{equation}
P^\Delta (U) = \left\{\begin{matrix}
P_0(U)\frac{\tau_E}{\tau}              &  U > U_\Delta\\
P_0(U)\frac{U^2\,\tau_E}{U^2_\Delta\,\tau} & U \leqslant U_\Delta
\end{matrix}\right.\,,
\eqlab{PU-Epar}
\end{equation}
where the energy-loss time in the presence of an electric field is given by \eqref{tau-E} and
\begin{equation}
U_\Delta = \sqrt{\frac{(m_0^2+P_\perp^2)\,\tau_E}{2\Delta}}\,.
\end{equation}
In the absence of an electric field we have, of course, $\tau_E=\tau$ given by \eqref{tau}.

\begin{figure}[h]
        \centering
        \begin{subfigure}[h]{0.45\textwidth}
                \includegraphics[width=\textwidth]{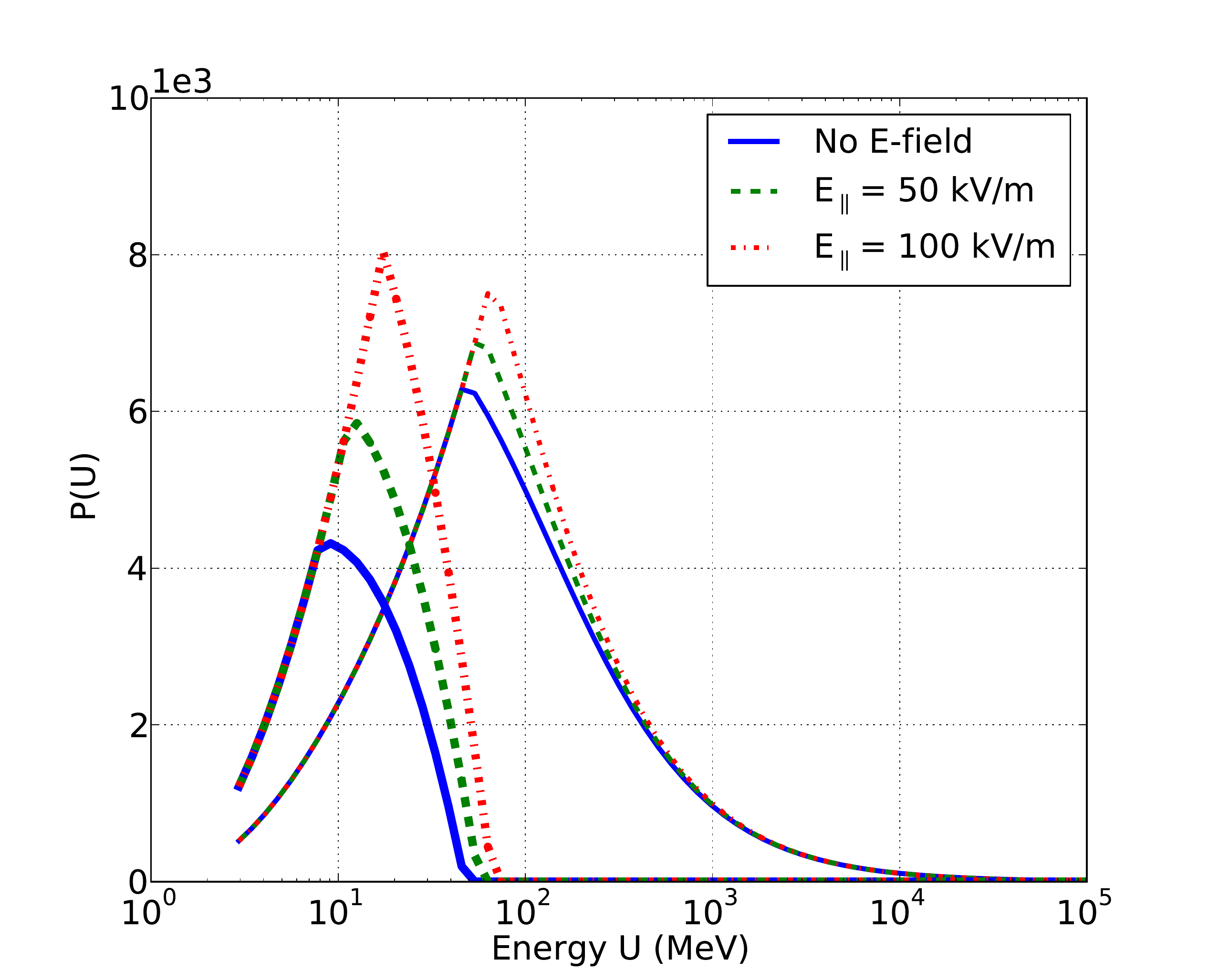}
               \caption{Analytical estimates} \figlab{E_spec_distance-a}
        \end{subfigure}
        \begin{subfigure}[h]{0.45\textwidth}
                \includegraphics[width=\textwidth]{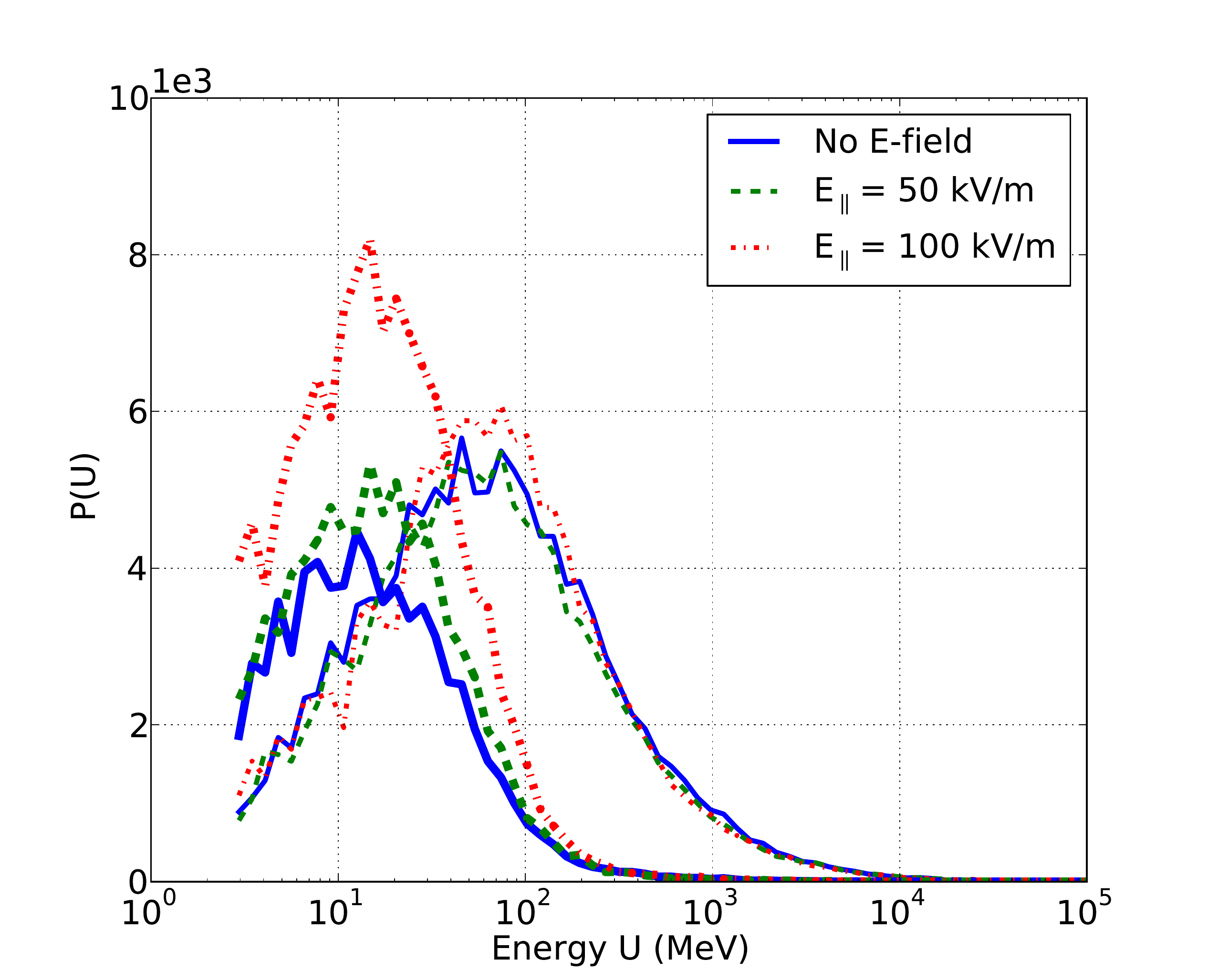}
                   \caption{CORSIKA simulations} \figlab{E_spec_distance-b}
        \end{subfigure}
      \caption{The energy spectrum of electrons within a distance of 3~m (thin curves) and from 3~m to 10~m (thick curves) behind the shower front for  $E = 0$, ${E_\parallel} = 50$ kV/m and ${E_\parallel} = 100$ kV/m at the shower maximum from analytical calculations (top panel) and from full CORSIKA simulations (bottom panel).}
\figlab{E_spec_distance}
\end{figure}

It should be noted that the factor in \eqref{PU-Epar},
\begin{equation}
\frac{U^2\,\tau_E}{U^2_\Delta\,\tau}=\frac{2 U^2\Delta}{(m_0^2+P_\perp^2)\tau} \;,
\end{equation}
is independent of $\tau_E$. At energies below $U_\Delta$ (which itself depends on the magnitude of the electric field), the number of electrons within a certain trailing distance is independent of the electric field. This feature is seen in the estimates of \figref{E_spec_distance-a} as well as in the full Monte Carlo simulations of \figref{E_spec_distance-b}. At energies exceeding $U_\Delta$ the number of electrons is proportional to $\tau_E$ and thus increases with the strength of the electric field. This increase is however very moderate compared to the increase of the number of electrons at larger distance, see \figref{E_spec_distance}.

\begin{figure}[h]
                \includegraphics[width=0.48\textwidth]{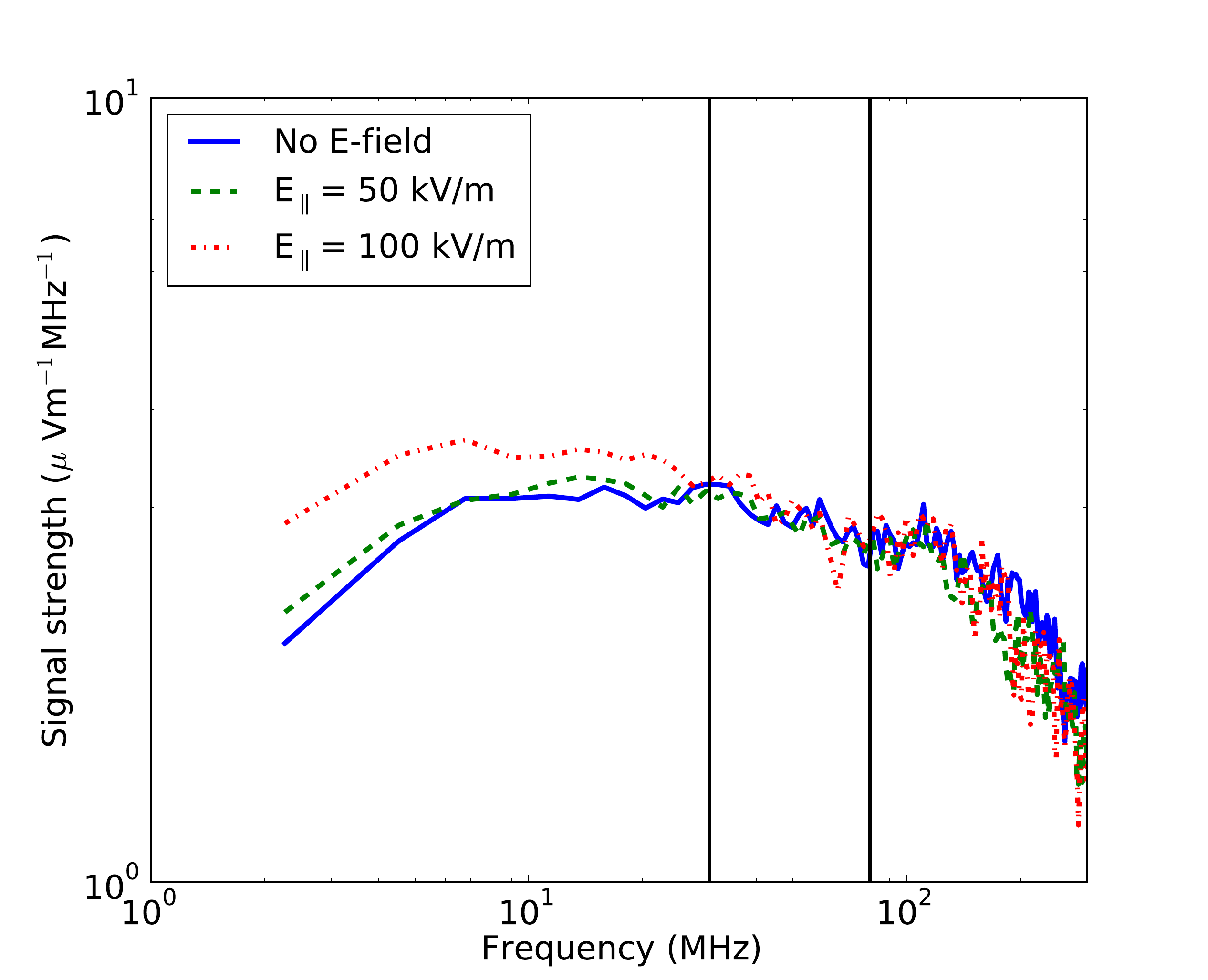}
               \caption{The signal strength, as obtained from CoREAS simulations, as function of frequency for a vertical shower of $10^{15}$~eV at 50~m from the core in the absence and in the presence of parallel electric fields. The black vertical lines represents the LOFAR LBA frequency window.
               }
             \figlab{power_freqdomain}
\end{figure}
In the distance interval from 3~m to 10~m behind the shower front, the number of electrons with an energy less than 20~MeV are significantly enhanced. As a result, there is a strong increase in coherent radiation at larger wavelengths, well below the frequency range of LOFAR LBA as it is shown in \figref{power_freqdomain}.

We can thus conclude that an accelerating electric-field parallel to the shower axis increases the total number of electrons. The enhancement in the number of electrons occurs mainly at low energy and thus their relative velocity with respect to the shower front is larger than for the high-energy electrons. As a consequence, they are trailing much more than 3~m behind the shower front. Their radiation is thus not added coherently in the LOFAR frequency range but instead in the frequency below 10~MHz. Therefore, the effects of parallel electric fields cannot be observed by LOFAR operating in the frequency of 30-80~MHz. These effects should be measurable at a lower frequency range.

\subsection{Influence of $\bf{E_\perp}$}\seclab{E-perp}

One important result from the CORSIKA simulations, shown in \figref{mean_distance_Eperp}, is that the median trailing distance behind the shower front increases rapidly with increasing transverse force working on the electrons. In order to get some more insight into the dynamics we try to reproduce this in our simplified picture.

\begin{figure}[h]
                 \centering
                \includegraphics[width=0.45\textwidth]{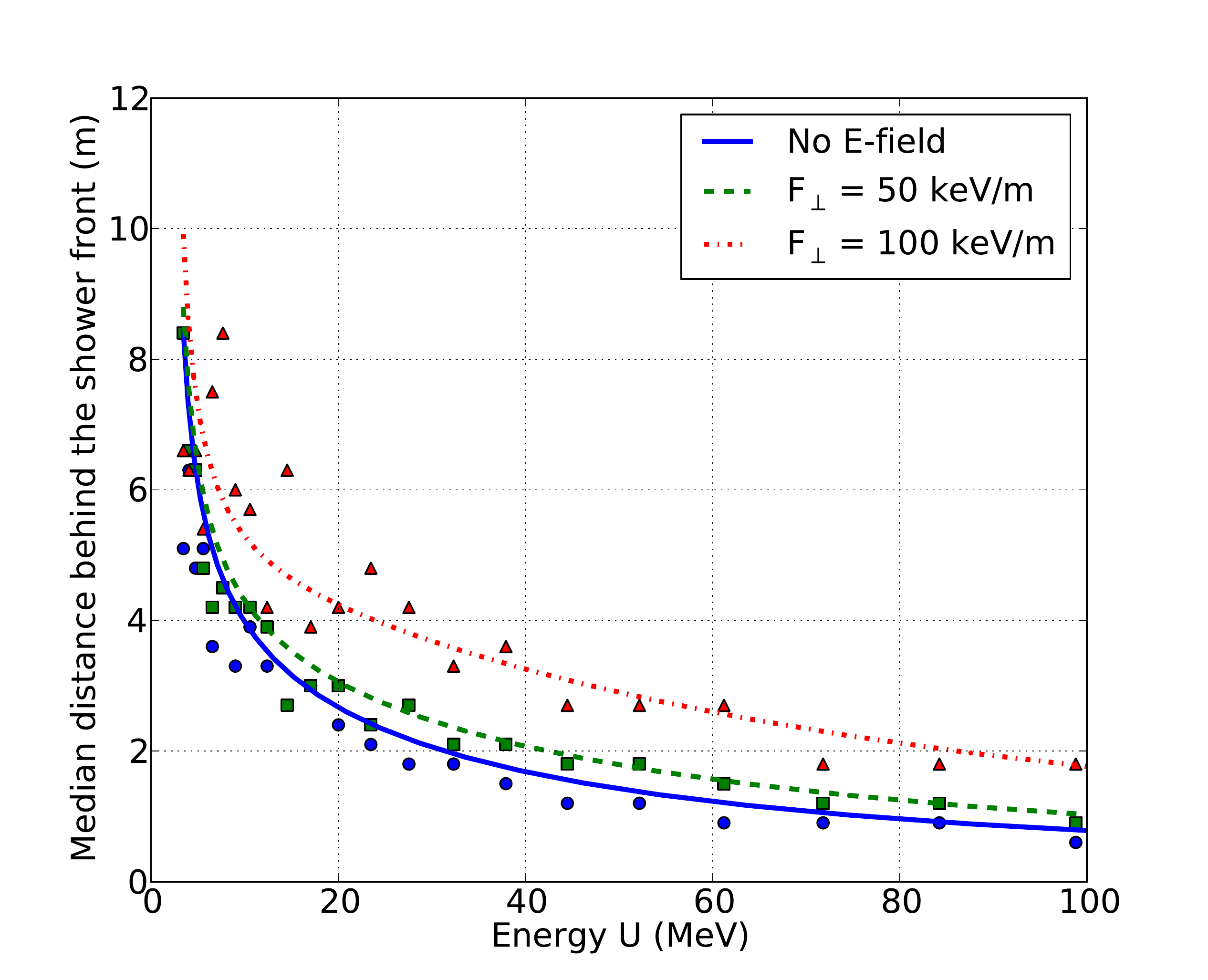}
               \caption{The median distance by which the electrons trail behind the shower front as function of their energy in the absence and in the presence of a  perpendicular electric field as obtained from CORSIKA simulations (markers) in compared to the model expectation (solid and dashed curves).
               }
            \figlab{mean_distance_Eperp}
\end{figure}

When an electric field perpendicular to the shower axis is applied, there is a transverse electric force acting on the electrons and positrons. The transverse net force which is the vector sum of the transverse electric force and the Lorentz force,
\begin{equation}
\bf{F}_\perp = q\left(\bf{E}_\perp+\bf{v}\times\bf{B}\right) \;,
\end{equation}
causes the electrons and positrons to move in opposite directions. Since the field is perpendicular to the main component of the velocity, no appreciable amount of work is done and the electron energy is not really affected. Thus, the energy-loss time of the electrons remains almost unchanged. As a result the perpendicular electric field does not change the total number of electrons given by \eqref{NU} in complete accordance with the results of Monte-Carlo simulations.

The electrons are subjected to the transverse force $\bf{F}_\perp$ giving rise to a change in transverse momentum
\begin{equation}
F_\perp = \frac{dP_\perp}{dt}\,.
\end{equation}
The average initial transverse momentum in the direction of the force vanishes. The mean transverse momentum is thus
\begin{equation}
\bar{P}_\perp(t) = F_\perp\,t\;,
\end{equation}
where $t$ is the time lapse after creation. Due to the action of the force, the electrons drift with a velocity
\begin{equation}
\bar{v}_\perp(t) = \frac{\bar{P}_\perp}{U}=\frac{F_\perp\,t}{U} \;,
\eqlab{v_perp}
\end{equation}
where $U$ is the energy of the electrons. The random component of the transverse momentum is taken into account in the effective transverse mass as introduced in \eqref{PPerpAve}. The parallel velocity is thus
\begin{equation}
v_\parallel(t) ={P_\parallel \over U}=\sqrt{1 - {m_\perp^2+\bar{P}_\perp^2(t)\over U^2} } \approx 1-\frac{1}{2}\frac{m_\perp^2+\bar{P}_\perp^2(t)}{U^2}\;.
\label{beta_para}
\end{equation}
The transverse velocity increases when the net force increases, the longitudinal velocity reduces because the total velocity cannot exceed the light velocity. Since the transverse velocity is small, even in strong fields, the electrons trailing within 3~m behind the shower front do not drift more than 100~m sideways. The distance of 100~m we had imposed in order to avoid corrections due to the curved shower front.
After a time $t$, the electrons are trailing behind the shower front by a distance
\begin{eqnarray}
l(t) &=& \int_{0}^{t}\left(c-v_\parallel\right)dt =  \int_{0}^{t}\frac{m_\perp^2+\bar{P}_\perp^2(t)}{2\,U^2}dt
\\ \nonumber
&=& {m_\perp^2 \,t\over 2 \, U^2} + \frac{F^2_\perp\,t^3}{6U^2}\,.\eqlab{time_d}
\end{eqnarray}
The median distance by which an electron can trail behind the shower front within its energy-loss time $\tau$ (see \eqref{tau}) is given by
\begin{equation}
D(E_\perp) = l(\tau/2)=\frac{\tau}{4 \, U^2}\left(m_\perp^2 + {1\over 12}F^2_\perp\,\tau^2\right) \;.
\eqlab{d-ave}
\end{equation}
This equation shows that $D$ quickly decreases with increasing energy as is seen in \figref{mean_distance_Eperp} as obtained from a CORSIKA simulation. With increasing $F_\perp$ the second term in \eqref{d-ave} increases quadratically giving rise to a rapidly increasing trailing distance as is also seen in the simulation. The median distance of low-energy electrons in the presence of the net force of 100~keV/m given by the simulations are larger than the simple prediction because the electron density is small at this energy range and as a sequence there is a fluctuation in the median distance.

\begin{figure}[h]
        \centering
        \begin{subfigure}[h]{0.45\textwidth}
                \includegraphics[width=\textwidth]{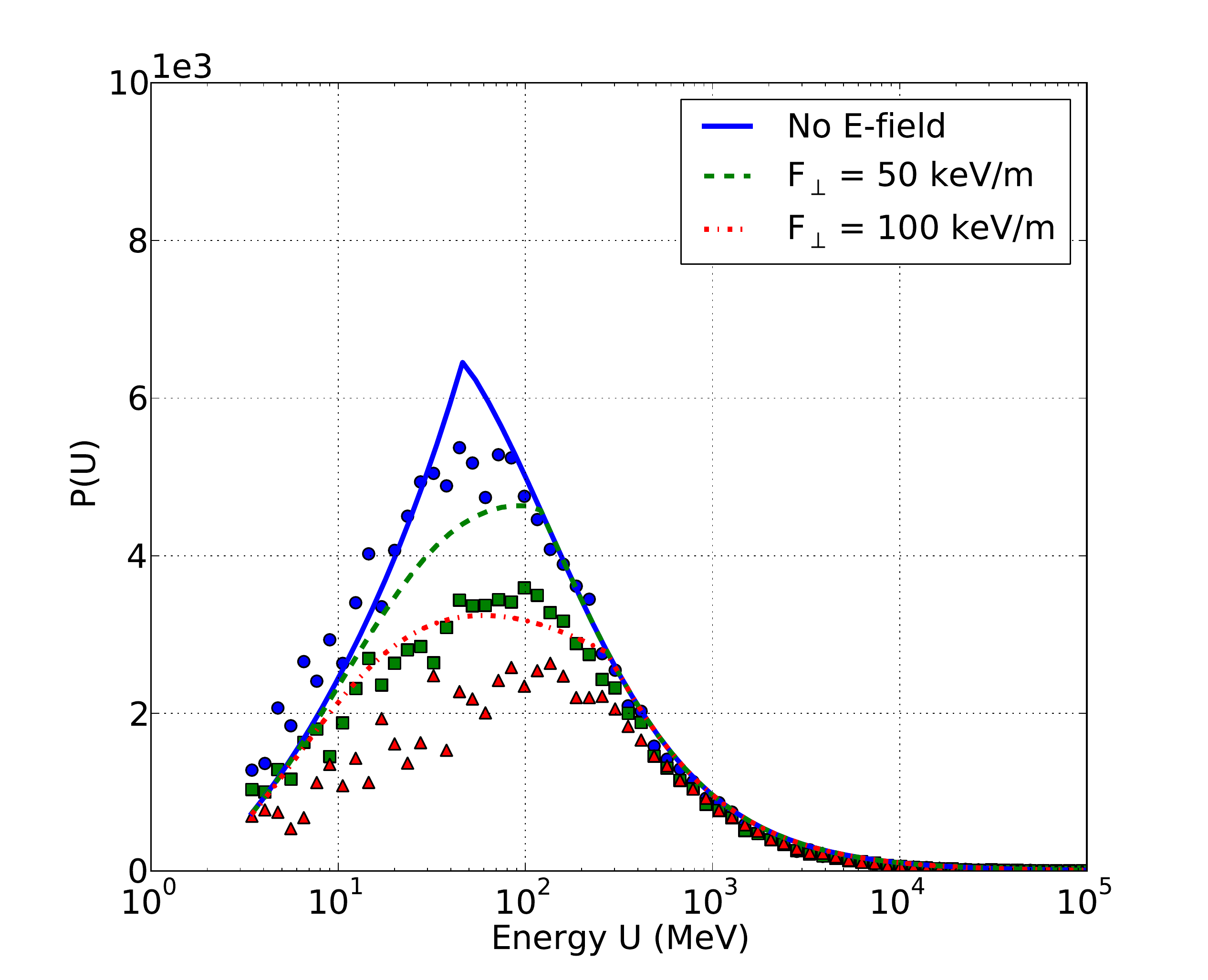}
        \end{subfigure}
    \caption{The number of electrons as function of their energy that are within 3~m behind the shower front with and without perpendicular electric fields at $X = 500$ g/cm$^2$ as calculated from \eqref{E_spec} (solid and dashed curves) and from the full CORSIKA simulations (markers).}
   \figlab{energy_spec}
\end{figure}
Essential for radio emission in the LOFAR LBA frequency band is the total number of the electrons within 3~m behind the shower front. The results of the CORSIKA simulation are displayed in \figref{energy_spec}. This shows that this number reduces as the transverse net force increases and that this decrease is strongest at those energies where the number of particles is maximal. In our simple picture the number of electrons within the distance $\Delta$ can be calculated by
\begin{equation}
{\cal P}^\Delta(U)=\int_{0}^{t_\Delta\leqslant \tau}\frac{P_0(U)}{\tau}dt=\left\{\begin{matrix}
P_0(U) & t_\Delta >  \tau\\
P_0(U)\frac{t_\Delta}{\tau} & t_\Delta\leqslant \tau
\end{matrix}\right.\,,
\eqlab{E_spec}
\end{equation}
where $t_\Delta$ is the root of \eqref{time_d} for $d = \Delta = 3$~m. Note that the equation gives one real root and two complex roots where the real root is taken since $t_\Delta$ is a physical quantity. \eqref{E_spec} can be simplified by introducing $\tau^m_\Delta=\min(t_\Delta,\tau)$ to $P^\Delta(U)=P_0(U)\tau^m_\Delta/\tau$. In \figref{energy_spec} the results from \eqref{E_spec} as well as the results from the the simulations are shown for different strength of electric fields. The number of low-energy electrons reduces in strong fields because of the increase in trailing behind while the number of higher-energy electrons is stable because they are still close to the front.

\begin{figure*}
        \centering
        \begin{subfigure}[h]{0.45\textwidth}
                \includegraphics[width=\textwidth]{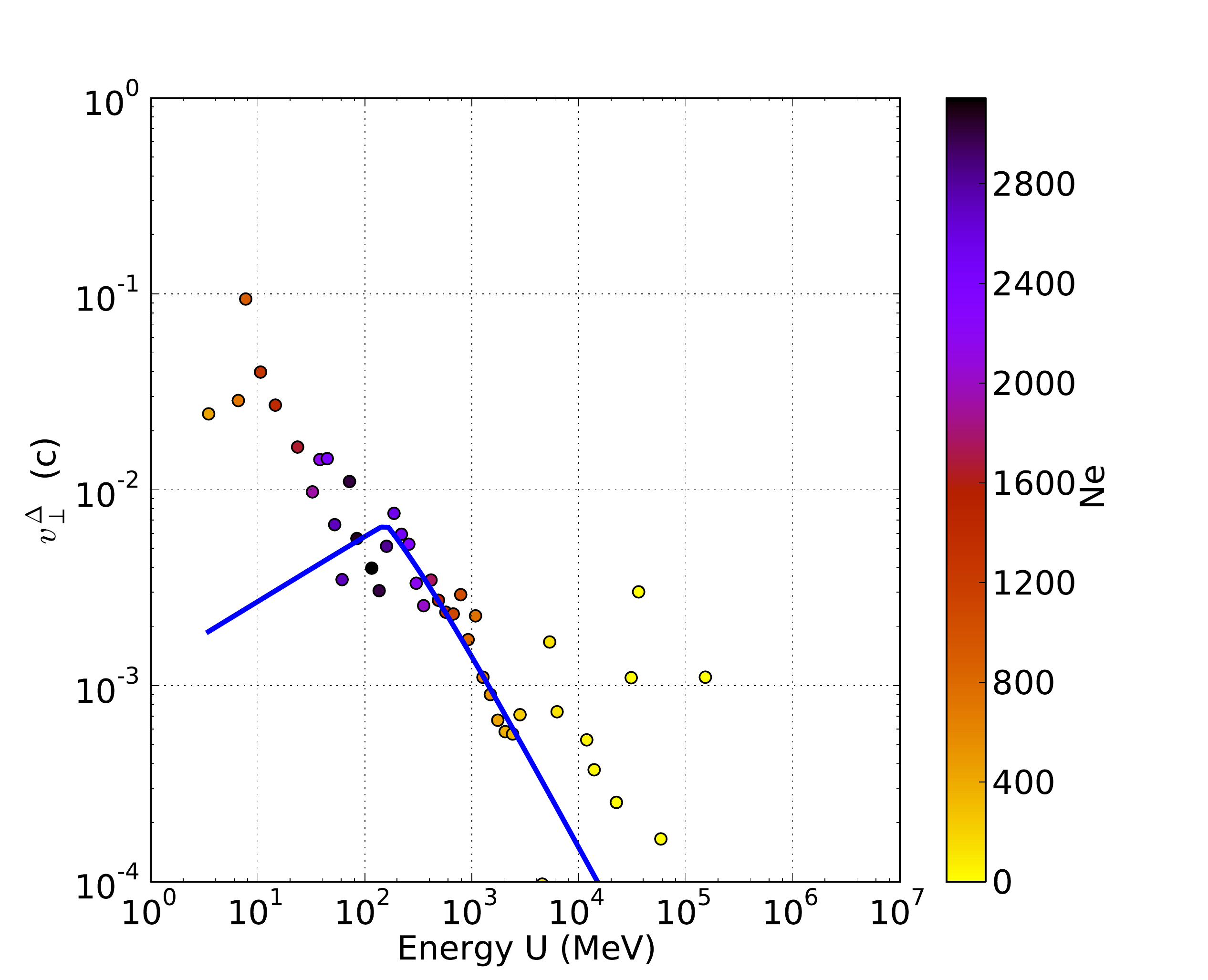}
               \caption{$F_\perp$ = 5 keV/m, $\Delta$ = 1~m}
        \end{subfigure}
        \begin{subfigure}[h]{0.45\textwidth}
                \includegraphics[width=\textwidth]{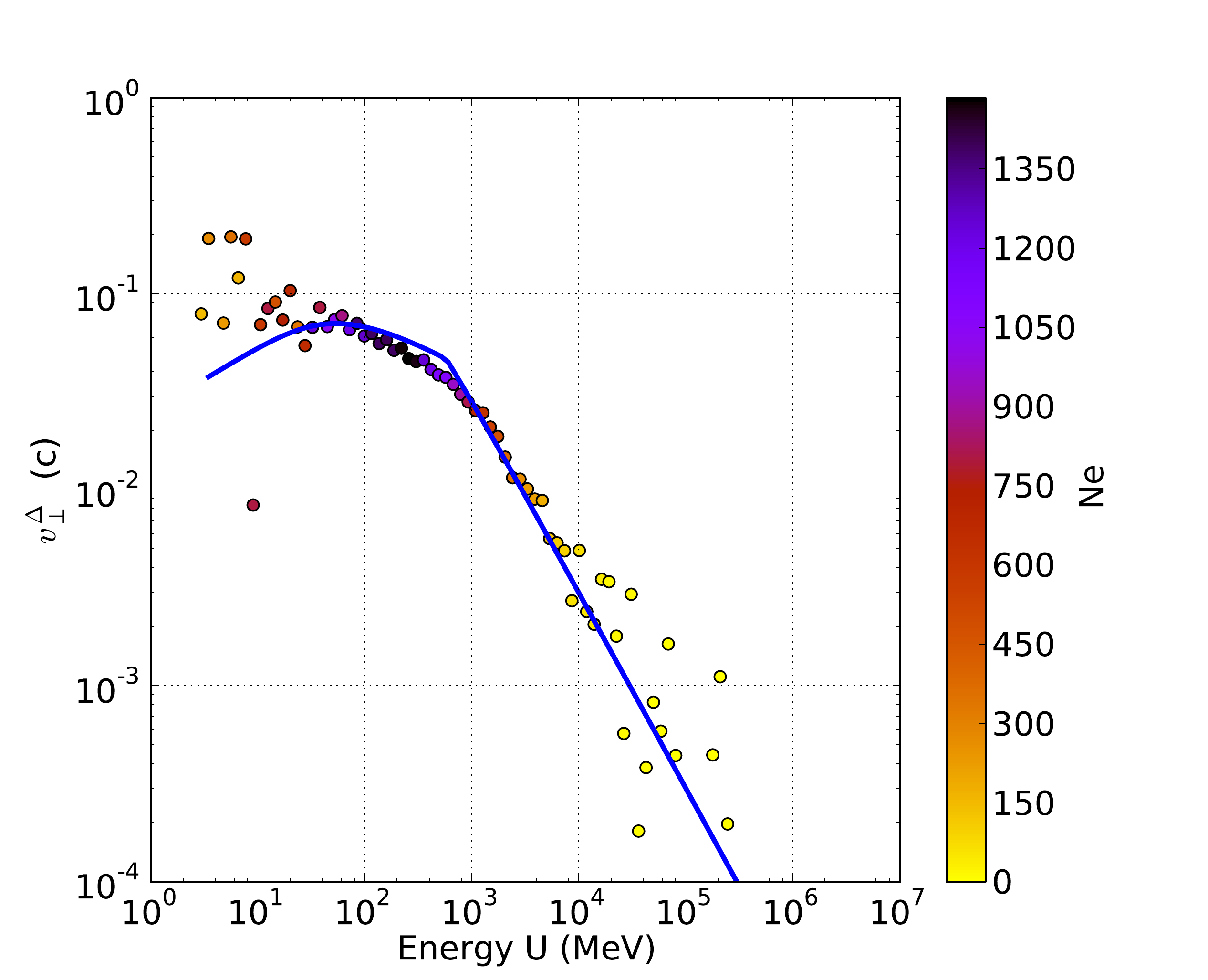}
                \caption{$F_\perp$ = 100 keV/m, $\Delta$ = 1~m}
        \end{subfigure}
       \begin{subfigure}[h]{0.45\textwidth}
                \includegraphics[width=\textwidth]{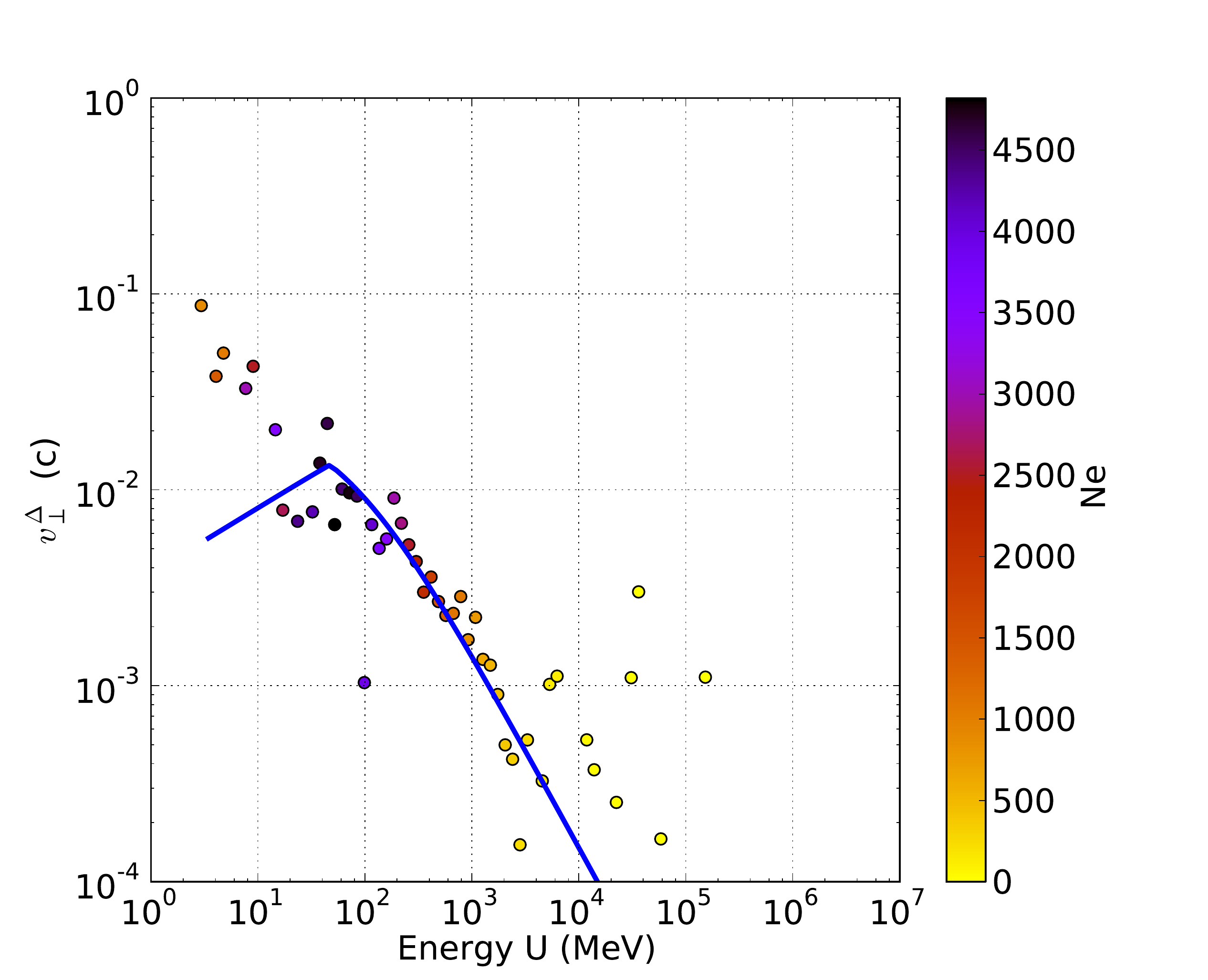}
                \caption{$F_\perp$ = 5 keV/m, $\Delta$ = 3~m}
        \end{subfigure}
        \begin{subfigure}[h]{0.45\textwidth}
                \includegraphics[width=\textwidth]{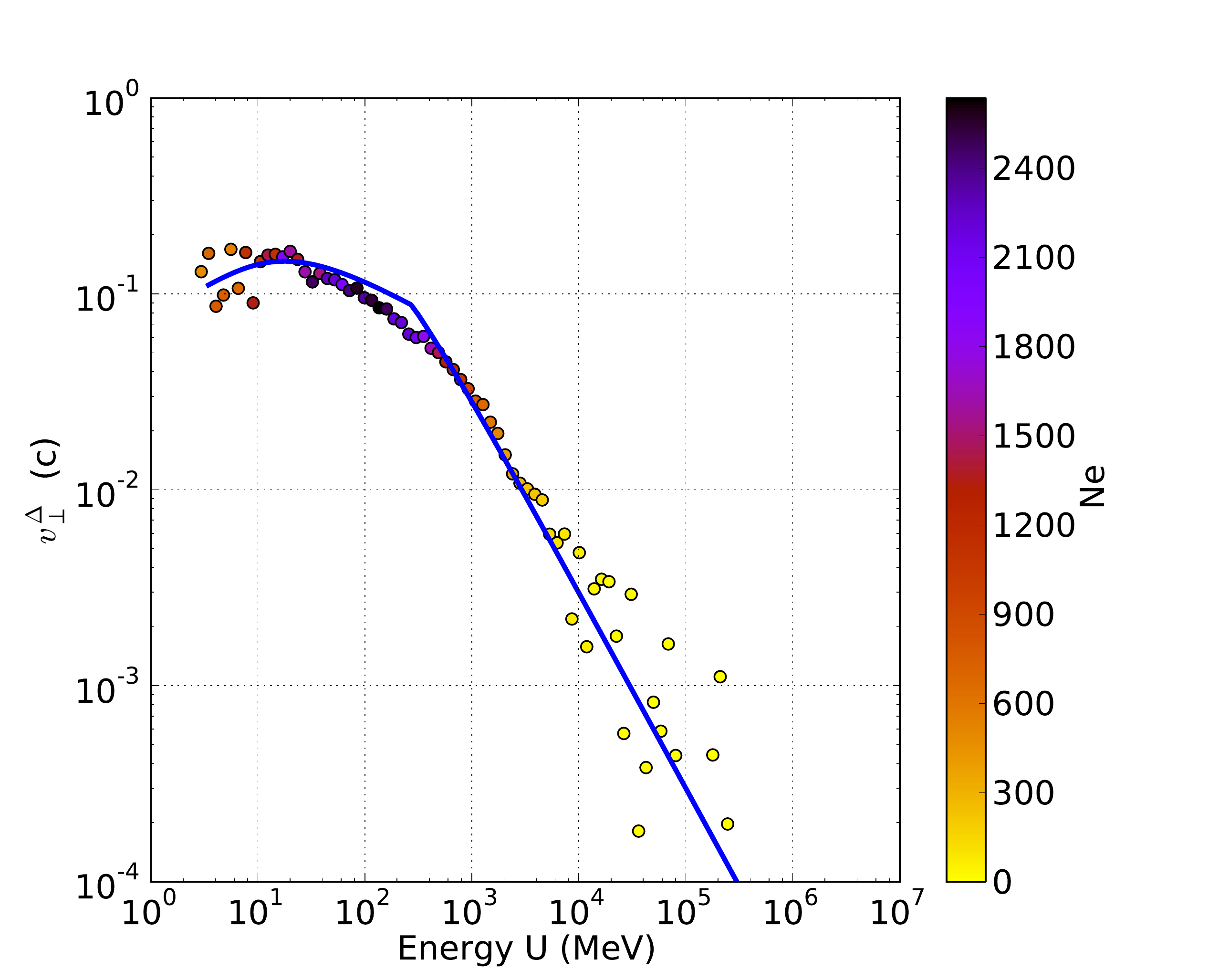}
               \caption{$F_\perp$ = 100 keV/m, $\Delta$ = 3~m}
        \end{subfigure}
       \begin{subfigure}[h]{0.45\textwidth}
                \includegraphics[width=\textwidth]{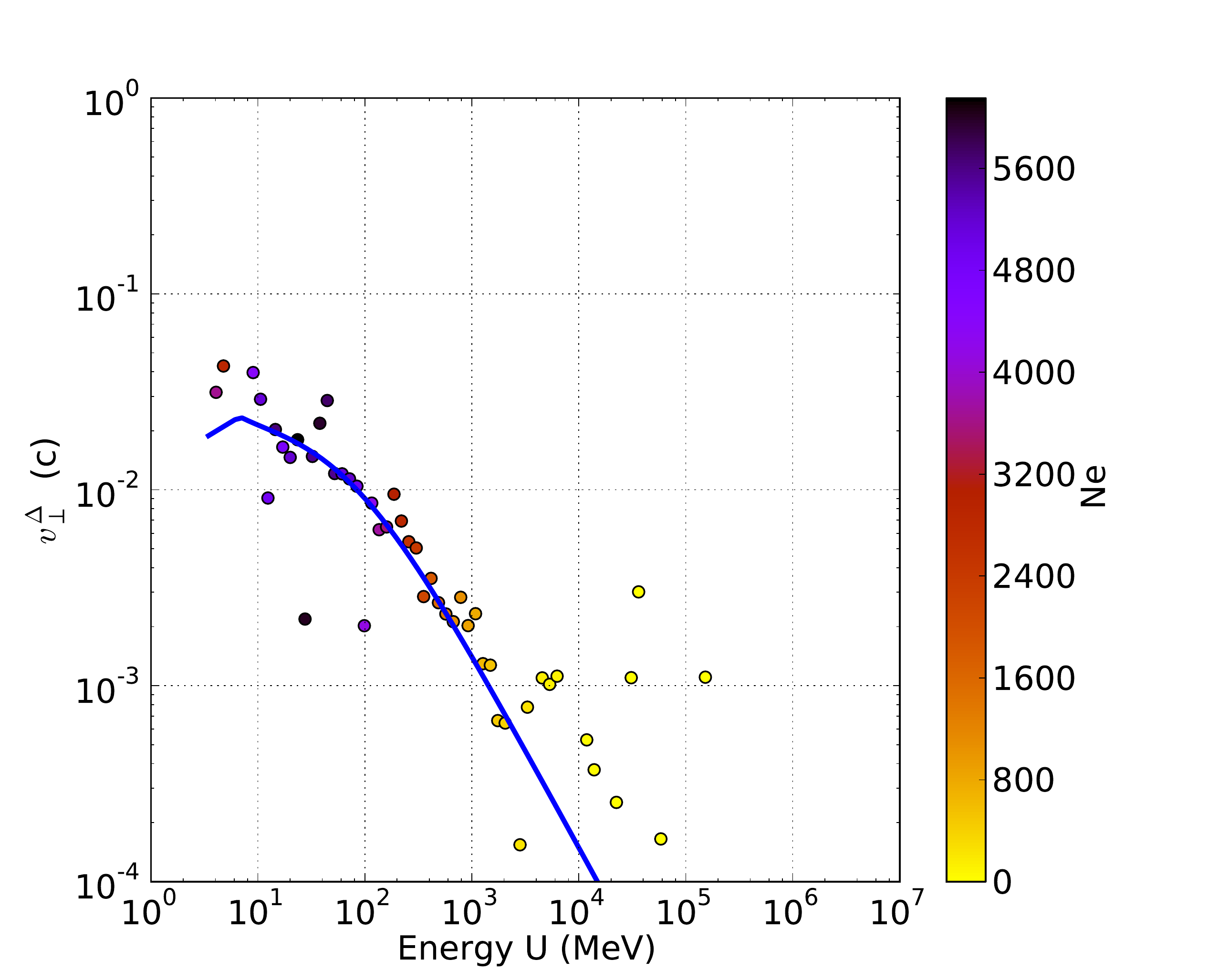}
                 \caption{$F_\perp$ = 5 keV/m, $\Delta$ = 10~m}
        \end{subfigure}
        \begin{subfigure}[h]{0.45\textwidth}
                \includegraphics[width=\textwidth]{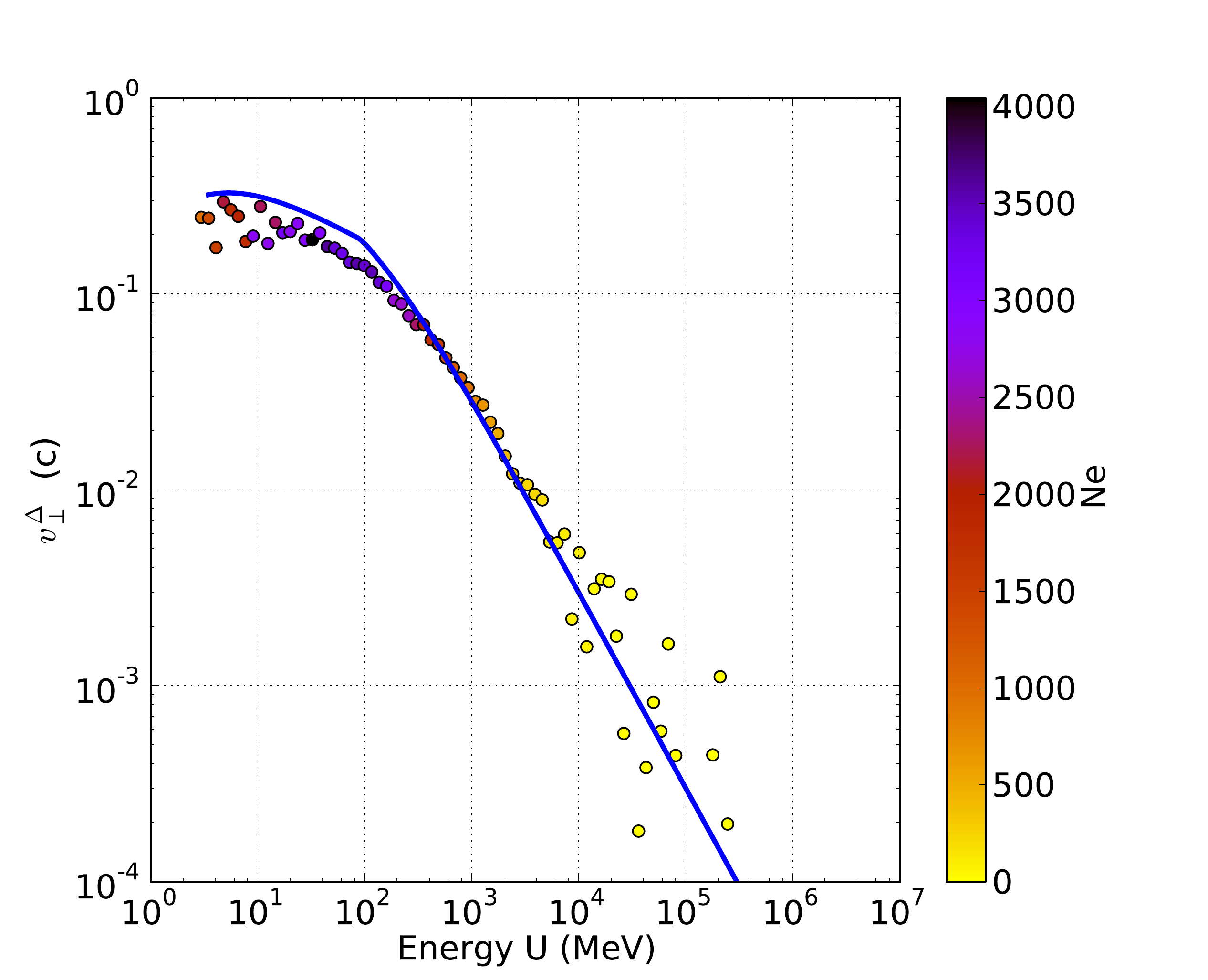}
                \caption{$F_\perp$ = 100 keV/m, $\Delta$ = 10~m}
        \end{subfigure}
    \caption{The drift velocity in the unit of the speed of light at $X_{max}$ of the electrons lagging within a distance $\Delta$ behind the shower front for different tranverse net forces from analytical estimates (blue curves) and from CORSIKA simulations results (markers). The colors of the dots represent the number of electrons.}
\figlab{vd_Eperp}
\end{figure*}

An important factor for the current is the drift velocity of the electrons. The results from the CORSIKA simulations is shown in \figref{vd_Eperp}. It increases with the strength of the net-transverse force and the distance behind the shower front.
Within our simple picture the mean drift velocity of electrons lagging within a distance $\Delta$ behind the shower front of the electrons is
\begin{equation}
v^\Delta_\perp(U) = {1\over t_\Delta^m} \int_0^{t_\Delta^m} v_\perp\,dt = {F_\perp\,\tau_\Delta^m\over 2\,U} \;,\eqlab{vd-perp}
\end{equation}
where $v_\perp$ is given by \eqref{v_perp}.
From \eqref{vd-perp} it follows that the drift velocity increases with distance to the shower front as also seen in the Monte Carlo simulations.
The main energy region that matters here is that between 50 MeV and 1000 MeV since the particle density in this energy range is largest.
Outside of this energy range, the scatter of the simulation results is large since the electron density is small and one suffers from poor statistics.

Combining \eqref{vd-perp} with the expression for the particle density one obtains the induced current carried by the particles within a distance $\Delta$ behind the shower front
\begin{equation}
j^\Delta_\perp(U) = e\,{P_0(U)\over \tau} \int_0^{t_\Delta^m} v_\perp\,dt = {e\,P_0(U)\,F_\perp\,(\tau_\Delta^m)^2\over 2\,\tau\,U} \;.\eqlab{j-perp}
\end{equation}

\begin{figure*}[htb]
                \includegraphics[width=.8\textwidth]{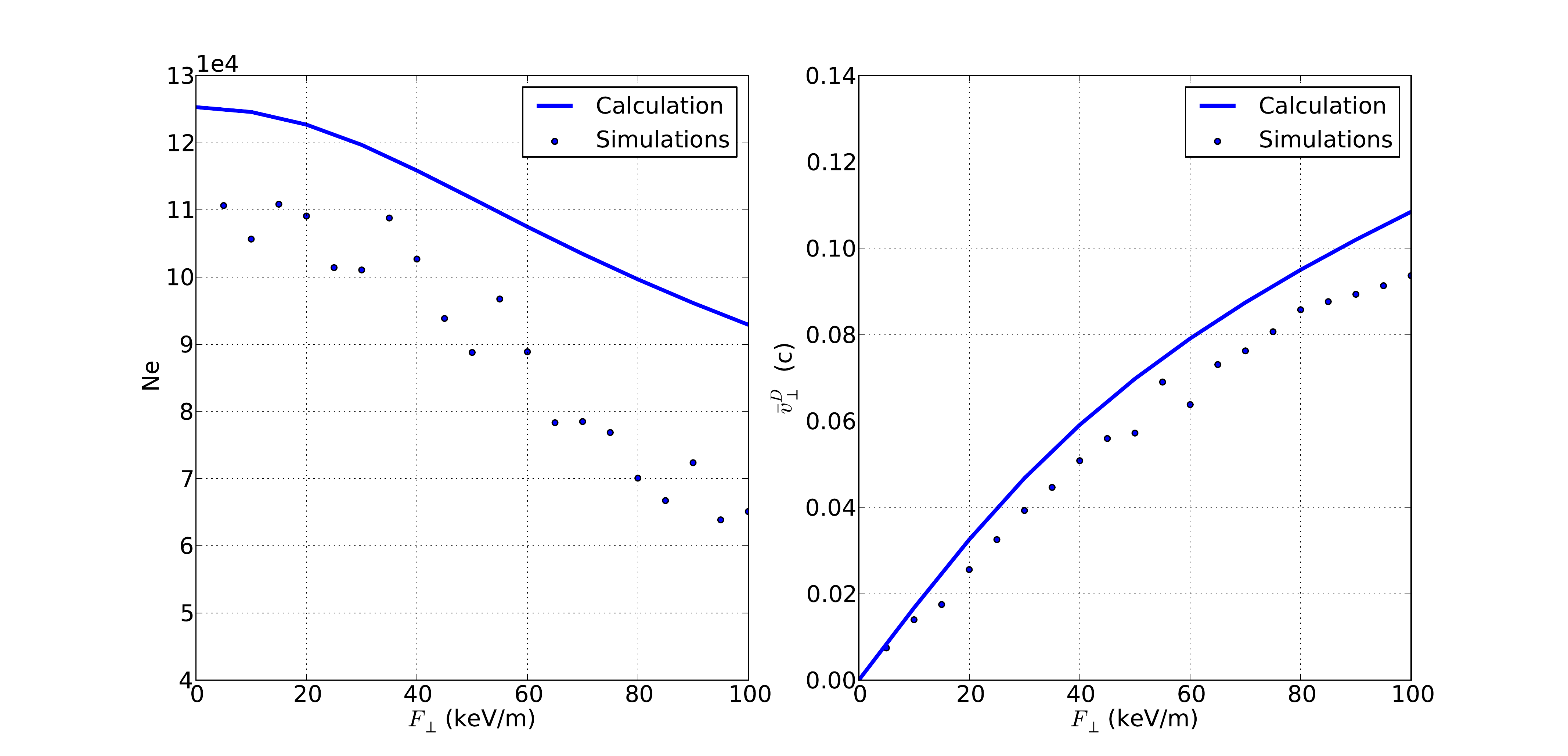}
               \caption{The number of electrons within 3~m behind the shower front (left panel) and their mean drift velocity (right panel) as a function of transverse net forces from analytical calculations and from CORSIKA simulations.}
\figlab{Ne_vd_Fperp}
\end{figure*}

\figref{Ne_vd_Fperp} displays the total number of electrons within 3~m behind the shower front and their mean drift velocity as a function of the net-transverse forces.
From the simple picture it is understood that in the presence of a perpendicular electric field the drift velocity of the electrons increases. However, they are lagging further behind the shower front. Therefore, the number of electrons within 3~m behind the shower front reduces, as shown in the left panel of \figref{Ne_vd_Fperp}.
The number in the simple picture is overestimated and the number does not drop as fast with increasing electric field as follows from the Monte Carlo calculation, but the trends match.

The simulation shows that the mean drift velocity increases with the net-transverse force (right panel of \figref{Ne_vd_Fperp}), however there is a change of slope at about 50~keV/m. In the simple picture this change of slope is due to the fact that the distance behind the front increases quadratically while the drift increases only linearly with increasing transverse force.
As a consequence, the induced electric current, which is the product of the number of electrons and their drift velocity, saturates at 50 keV/m and thus also the pulse amplitude since it is proportional to the current.

\begin{figure}[htb]
                \includegraphics[width=0.48\textwidth]{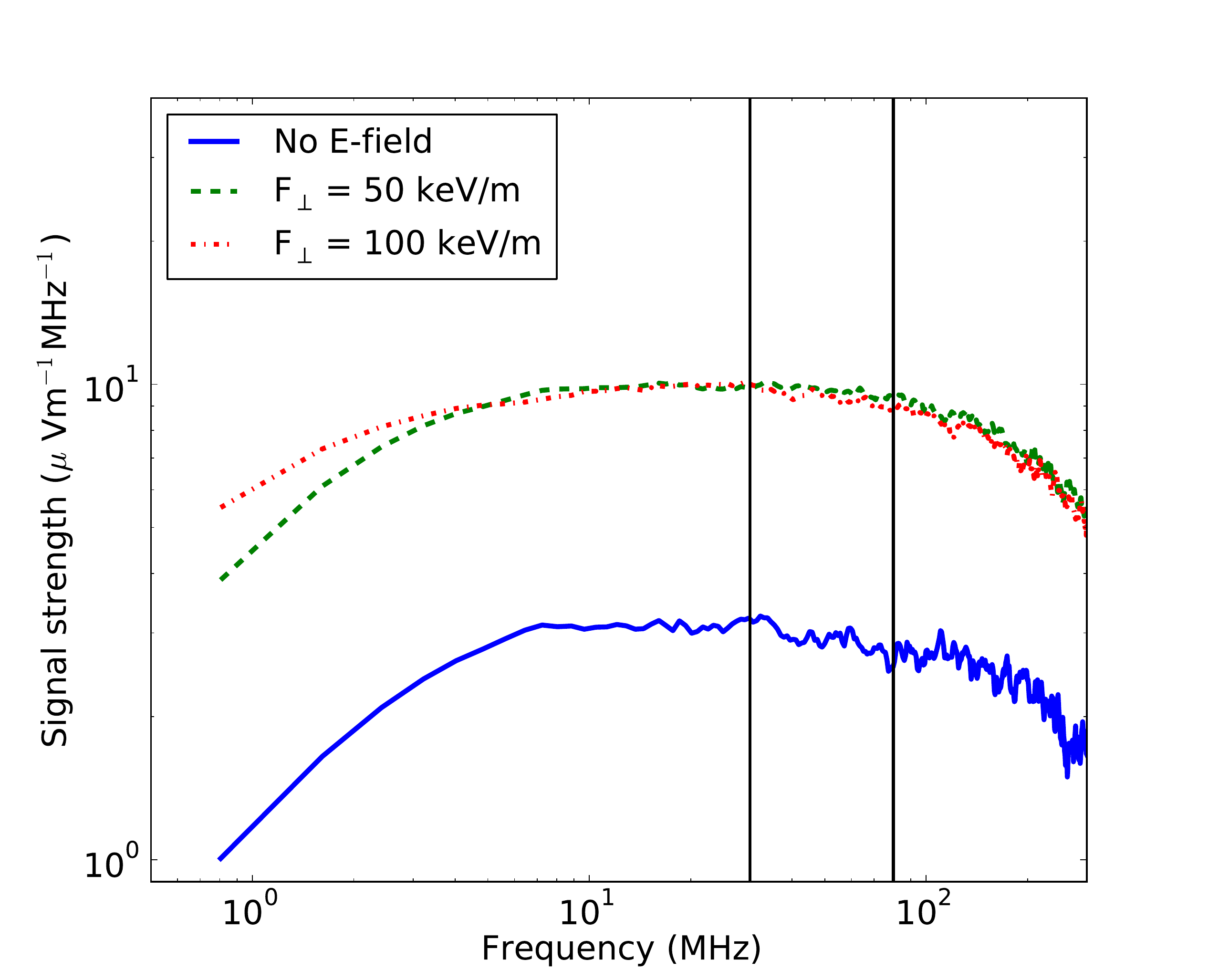}
               \caption{The same as displayed in \figref{power_freqdomain} for different strength of the net-transverse force.}
             \figlab{signalstrength_freqdomain_Fperp}
\end{figure}

Since, as just argued, for a large transverse force there is an increased drift velocity at larger distances behind the front one should thus expect an increased emission at longer wavelengths, well below the wavelength measured at LOFAR LBA. The effects of electric fields larger than 50 kV/m can be observed at lower frequency range as shown in \figref{signalstrength_freqdomain_Fperp}.

\subsection{Effects of electric fields in low-frequency domain}
As has been concluded in the previous two sections, the power as can be measured in the LOFAR frequency window of 30-80~MHz is strongly determined by the strength of the transverse electric field up to values of about 50~kV/m. In addition, parallel electric fields have small effects on the power in the frequency window 30-80~MHz. It was shown that at lower frequencies, the power keeps growing with increasing field strength up to at least 100 kV/m. In this section we investigate the usefulness of the 2- 9~MHz window in more detail. This frequency window is of particular interest for several reasons; i) it lies just below a commercial-frequency band, ii) the ionosphere shields the galactic background, and iii) the frequency is high-enough to observe a considerable pulse power.

\begin{figure}[htb]
        \centering
        \includegraphics[width=0.45\textwidth]{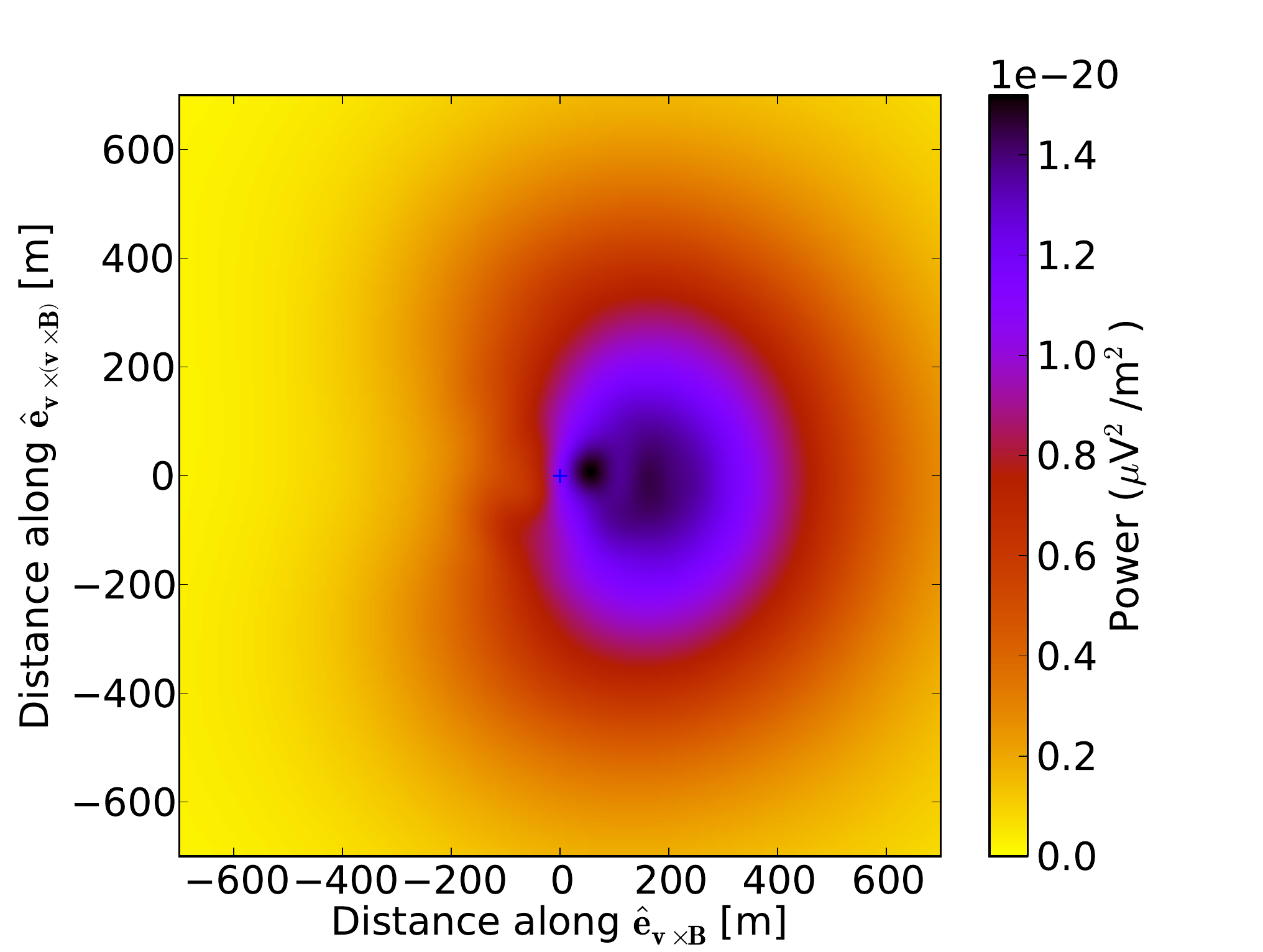}
        \includegraphics[width=0.45\textwidth]{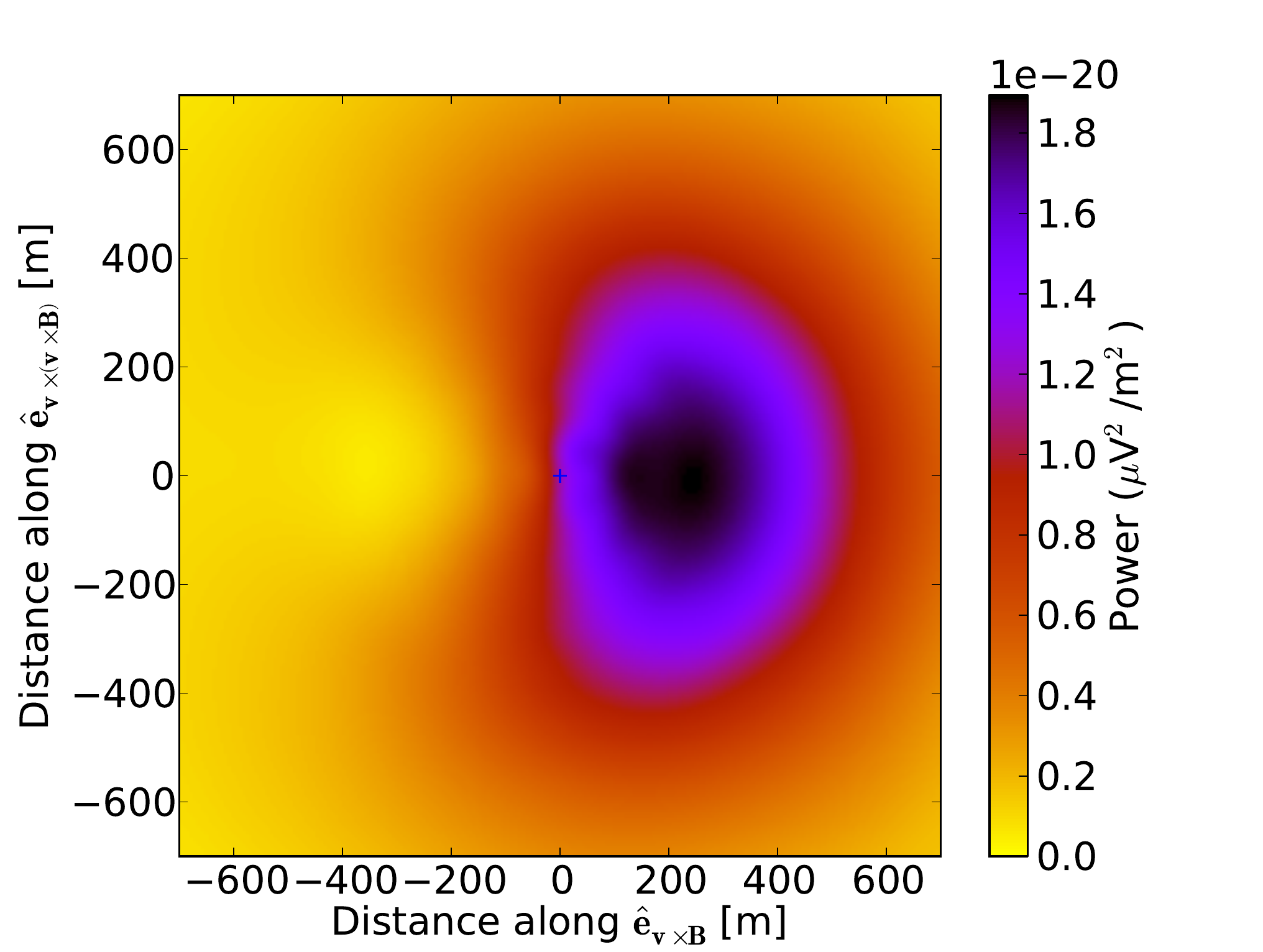}
        \includegraphics[width=0.45\textwidth]{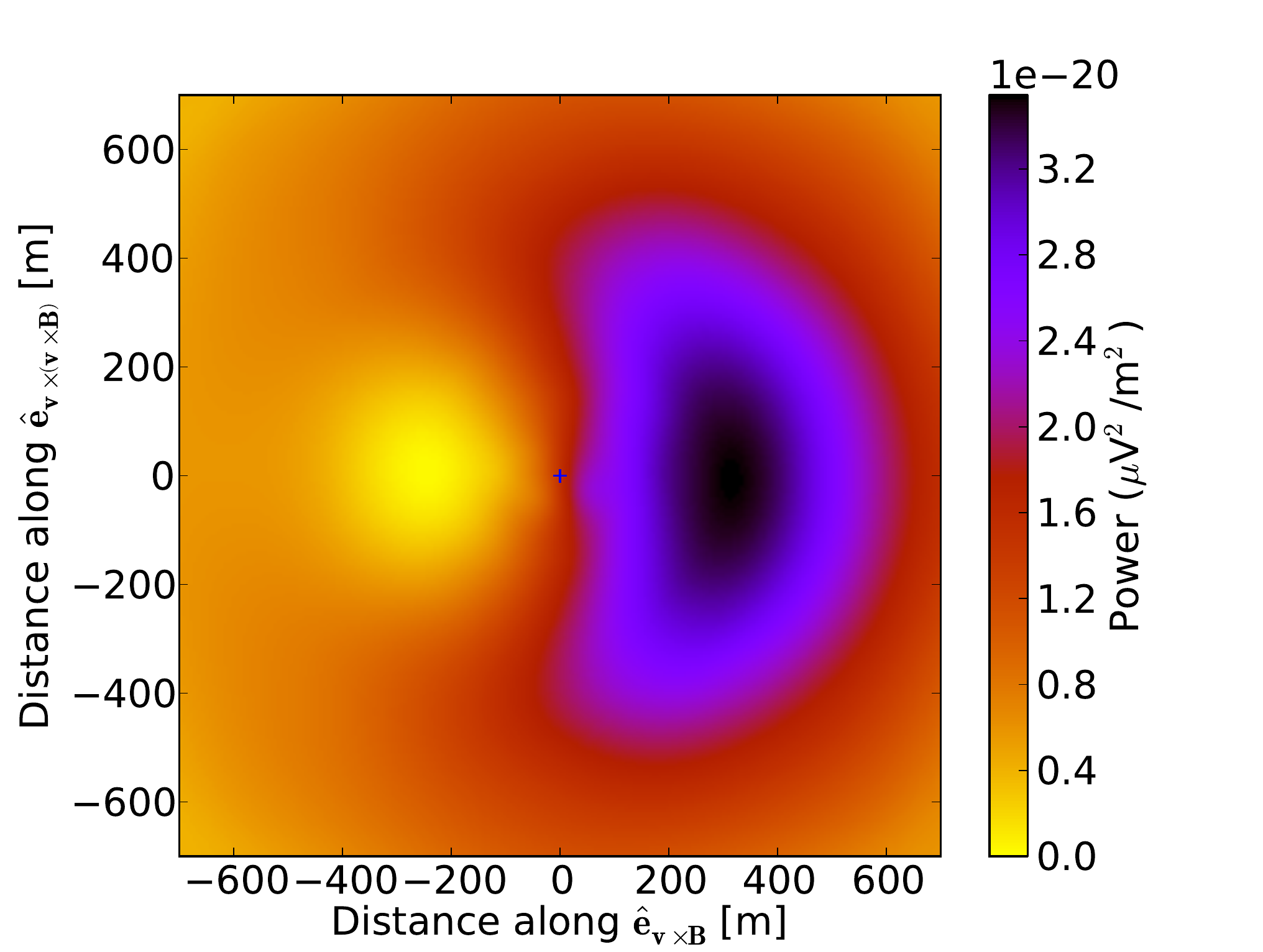}
    \caption{Intensity footprints of $10^{15}$~eV vertical showers for the  2 - 9~MHz band for the cases of no electric field (top), $E_\parallel$ = 50~kV/m (middle), and $E_\parallel$ = 100~kV/m (bottom).}
\figlab{Epara_lowf}
\end{figure}

\figref{Epara_lowf} shows that while parallel electric fields have only a minor effect on the emitted power in the frequency range from 30~MHz to 80~MHz, they have much larger effects on the power in the low frequency windows of 2--9~MHz. Here an increase of the peak-power with the strength of $E_\parallel$ is observed. Inside a strong parallel electric field, since the number of low-energy electrons increases and the number of low-energy positrons reduces, the charge-excess component becomes comparable to the transverse-current component. As a result of the interference the intensity at the highest electric field does not only have a strong maximum, but also a clear (local) minimum at 250~m from the shower core in the opposite direction as seen in the bottom panel of \figref{Epara_lowf}. Since these low-energy particles trail far behind the shower front, the change in the intensity pattern is not observed in the LOFAR-LBA frequency range.

\begin{figure}[htb]
        \centering
                \includegraphics[width=0.45\textwidth]{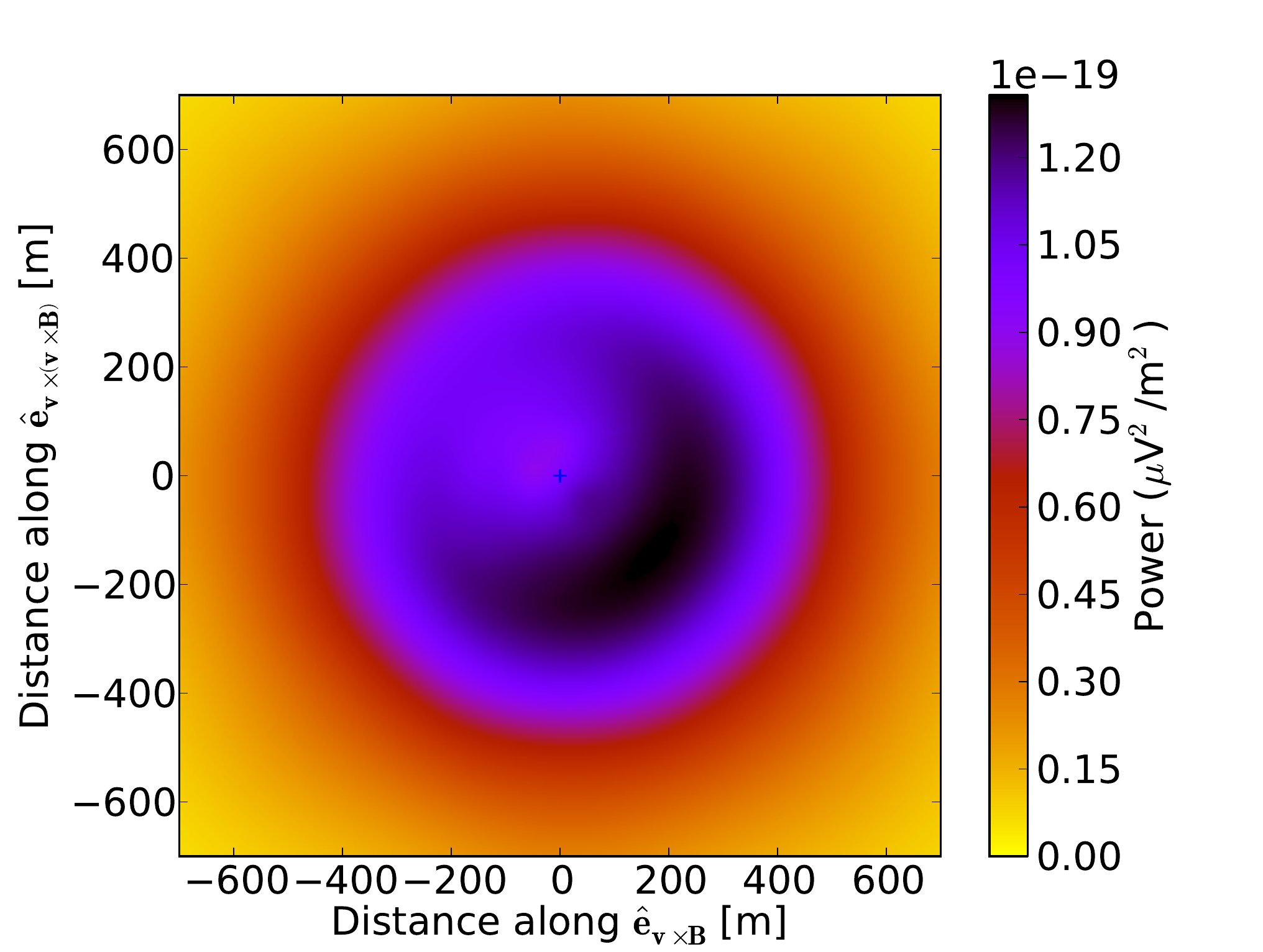}
                \includegraphics[width=0.45\textwidth]{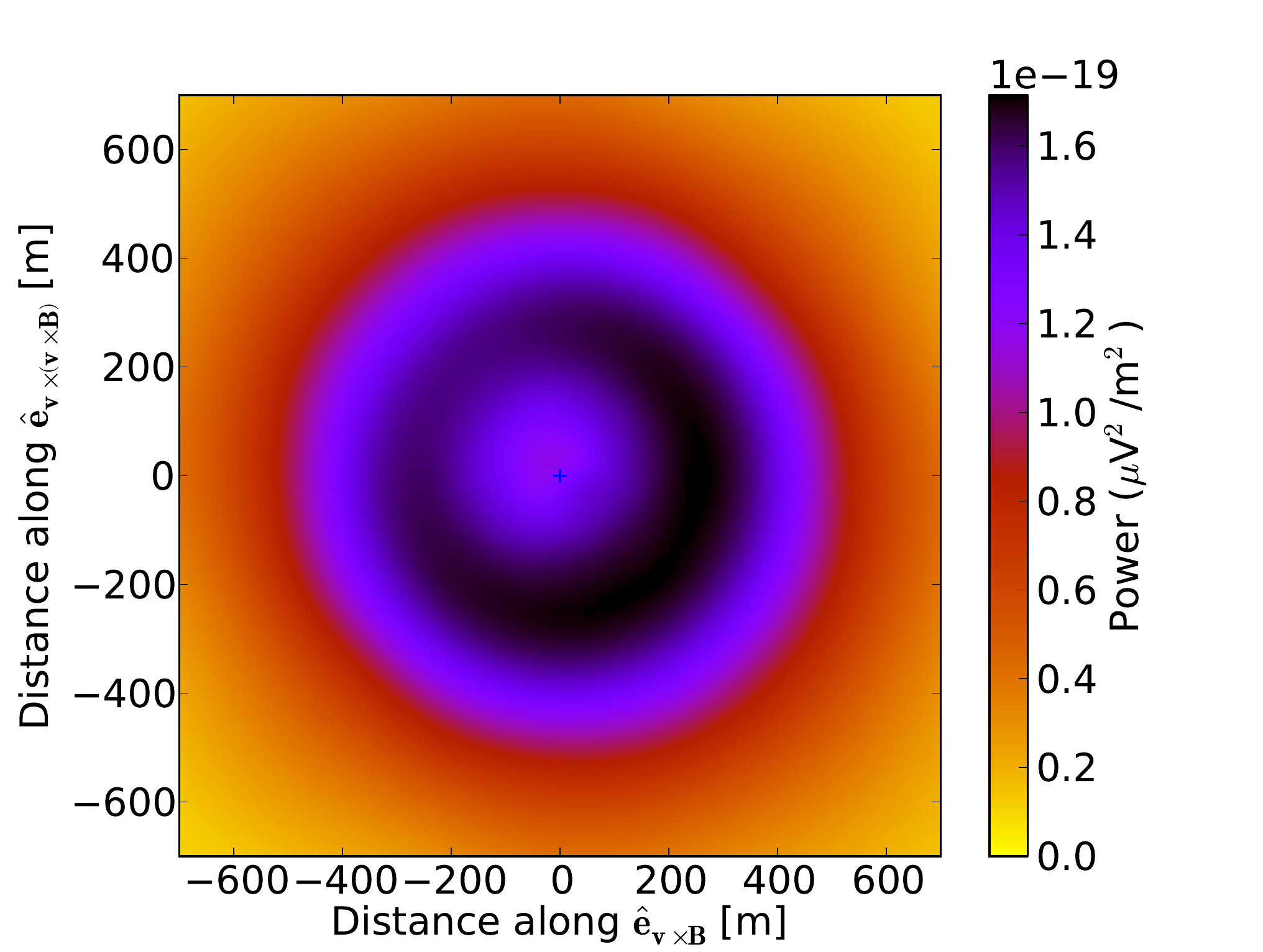}
    \caption{Intensity footprints of $10^{15}$~eV vertical showers for the  2 - 9~MHz band for the cases of $F_\perp$ = 50~keV/m (top) and $F_\perp$ = 100~keV/m (bottom).}
\figlab{Fperp_lowf}
\end{figure}

\figref{Fperp_lowf} displays that in the low frequencies window 2-9~MHz the maximum intensity increases with the net force to a very similar extent as in the frequency window 30-80~MHz.  The intensity footprint for the net force of 100~keV/m is more symmetric than the one for  50~keV/m because the electrons trail  further behind the shower front in a strong transverse electric fields and thus the charge-excess contribution becomes smaller.
The effects in the two frequency windows, 2-9~MHz and 30-80~MHz, are very similar although somewhat more pronounced at the lower frequencies. To have more leverage on the strength of the perpendicular component of the electric field one would need to go to even lower frequencies as is apparent from \figref{signalstrength_freqdomain_Fperp} which may be unrealistic for actual measurements.

The effects of parallel electric fields and large transverse electric fields are measurable at low frequencies from 2~MHz to 9~MHz. Since the intensity footprints become wider in the low-frequency domain (see \figref{Epara_lowf} and \figref{Fperp_lowf}), measuring signals in this range requires a less dense antenna array than in the LBA frequency domain.

\subsection{Adapting distance of the effects of E-fields}\seclab{adad}

\begin{figure}[htb]
                \includegraphics[width=0.48\textwidth]{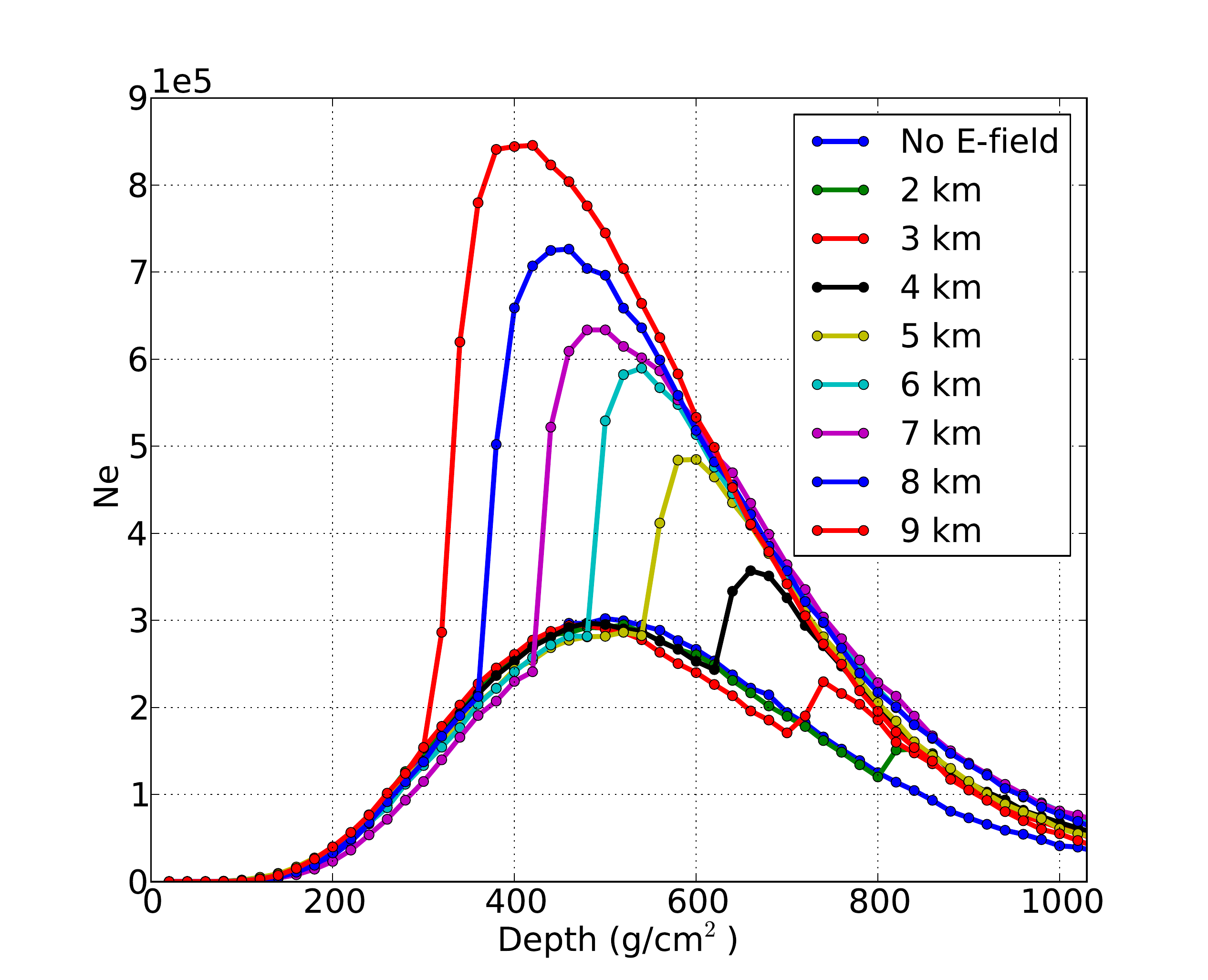}
               \caption{The number of electrons with kinetic energy larger than 401~keV as a function of atmospheric depth for 10$^{15}$ eV vertical showers, for electric fields as indicated at several altitudes. The depth of 0 means very high up in the atmosphere and the ground level is at the depth of 1036~g/cm$^2$.}
             \figlab{Ne_depth}
\end{figure}

One interesting aspect to study is the intrinsic distance along the track of the shower over which the electric fields are averaged using radio emission from air showers as a probe. It was also shown in Ref.~\cite{Buitink:2010} that the number of positrons adapts quickly to the expected number when the electric field is switched on. To study this in more details we have changed the magnitude of the electric field at a certain height and determined the number of particles as function of height in a CORSIKA calculation. The result is shown in \figref{Ne_depth} where the particle number is plotted as function of slant depth in stepsizes of 20 g/cm$^2$ for vertical showers. One remarkable feature one observes is that the particle number in the shower approaches a new equilibrium value which is apparently independent of the shower history. The distance over which this re-adjusting happens, the adapting distance, varies with height. It equals about 20 g/cm$^2$ at the height of 2 km and increases with the altitude to 80 g/cm$^2$ at 9 km.
Within our simple picture we expect this to vary as
\begin{equation}
X_a= \rho c \tau \;,
\end{equation}
where an appropriate averaging over particle energies should be performed. At large heights, where the shower is still young and dominated by high energy particles we expect on the basis of the energy loss times (see \figref{def_tau}) a longer distance of the order of $X_0$ shortening near the shower maximum where lower energetic electrons dominate.
Near the ground at LOFAR the atmospheric electric fields are small~\cite{Acounis:2012} and thus do not affect the number of particles in the air showers at ground-level. Therefore, the scintillators on the ground are not influenced by the atmospheric electric fields in clouds. This is supported by Ref.~\cite{Chilingarian:2014} (Figure 7) where it is shown (in mountain-top observations) that an enhanced rate of particles is only correlated to strong fields (fields in excess of 30~kV/m) at the height of the particle detectors.

\section{Conclusion}

We have studied in detail the effects of atmospheric electric fields on the structure of extensive air showers. In particular we have focussed on the distribution of the particles in the shower disk. The effects depend on the orientation of the field with respect to the plane of the disk. This is because in earlier work we observed that atmospheric electric fields strongly influence the radio emission from extensive air showers. The simulations showed that the intensity of the radiation is almost independent of the strength of the electric field parallel to the shower direction. We also observed a peculiar dependence of the intensity on the strength of the field perpendicular to the shower. This picture is supported by air-shower Monte Carlo simulations using the CORSIKA code.

In order to understand these dependencies we have performed Monte Carlo simulations of the dynamics of the electrons using the CORSIKA code. To understand the power as has been observed at LOFAR the number of particles and their drift velocities in a layer of about 3 meter (half the wavelength) behind the shower front is critical. The Monte Carlo simulations indicate a non-trivial dependence on the strength of the applied electric field.

To gain some more insight we have developed a simple picture where electrons are created at the shower front, move under the dynamics of the applied electric field, and disappear from the calculation after their energy loss exceeds a certain value. Under the influence of an accelerating electric field the energy-loss time of electrons increases which increases their numbers. However, at longer times after they have been created they will be at increasingly large distances behind the shower front and have moved outside the coherence region. The effect of an electric field perpendicular to the shower direction is more difficult to visualize since there are two counter-acting effects to consider. A perpendicular field will accelerate electrons in the transverse direction and thus have the effect to increase the current though the total number of electrons hardly increases. Since particles move relativistically an increased transverse velocity will result in a decrease of the longitudinal velocity since the total velocity cannot exceed the light velocity. As a result - for sufficiently large strength of the transverse electric field - they will trail further than the coherence length behind the shower front and thus not contribute to radio emission at the observed frequencies. The balance between these two effects results in an initial increase of the emitted intensity proportional to the applied field followed by a regime where the intensity is roughly constant when the field exceeds a critical value of around 50 kV/m at the altitude of 5.7~km.

In order to increase the sensitivity of the measurements to atmospheric electric fields it is shown that the deployment of antennas operating in the frequency window 2-9~MHz would be beneficial. This frequency interval is not subject to the galactic background because the ionosphere is not transparent at this frequency range.
The precision of the electric-field determination could also be increased by reducing the energy threshold of the measurement by decreasing the trigger threshold since this allows to observe more air showers and thus improving the sampling.
Such a study would not only deepen our understanding of the influence of atmospheric electric fields on air showers and their radio emission but also provide a powerful tool to study the electric fields in thunderclouds. The latter would be important to resolve the issue of lightning initiation~\cite{Dubinova:2015, Dwyer:2014, Gurevich:2013}.

\appendix
\section{CORSIKA}
In the CORSIKA simulations, we use the high-energy hadronic interaction model QGSJET-II~\cite{Ostapchenko:2006} and for the low-energy interactions we use FLUKA~\cite{Battistoni:2007}. Atmospheric electric fields are implemented by turning on the EFIELD~\cite{Buitink:2010} option in CORSIKA.
The geomagnetic field put in the simulations is the geomagnetic field at LOFAR.
The ``thinning" option with optimized weight limitation~\cite{Kobal:2001} is also used with a factor of 10$^{-6}$ to keep the computing times at a reasonable level.
We have simulated four kinds of iron showers: vertical showers of 10$^{15}$~eV, 10$^{16}$~eV, and 10$^{17}$~eV and inclined showers of 10$^{16}$~eV with a zenith angle of 30 degrees.
Of particular relevance for radio emission is $X_\mathrm{max}$ which is also subject to shower-to-shower fluctuations. To limit this effect, simulations are selected where $X_\mathrm{max}$ differs by not more than 30~g/cm$^2$.

\begin{acknowledgements}
The LOFAR cosmic ray key science project acknowledges funding from an Advanced Grant of the European Research Council (FP/2007-2013) / ERC Grant Agreement n. 227610. The project has also received funding from the European Research Council (ERC) under the European Union's Horizon 2020 research and innovation programme (grant agreement No 640130).   We furthermore acknowledge financial support from FOM, (FOM-project 12PR304) and the Flemish foundation of scientific research (FWO-12L3715N). AN is supported by the DFG (research fellowship NE 2031/1-1).

LOFAR, the Low Frequency Array designed and constructed by ASTRON, has facilities in several countries, that are owned by various parties (each with their own funding sources), and that are collectively operated by the International LOFAR Telescope foundation under a joint scientific policy.
\end{acknowledgements}
\bibliography{paper_bigthunderstorm}
\end{document}